\documentclass{aa} 

\usepackage{graphicx}
\usepackage{txfonts}
\usepackage{natbib}
\bibpunct{(}{)}{;}{a}{}{,}

\newcommand{\be}{\begin{equation}}
\newcommand{\ee}{\end{equation}}
\newcommand{\angstrom}{\mbox{\normalfont\AA}}

\def\apjl{ApJL}
\def\apjs{ApJS}

\newcommand{\Mpc}{$h^{-1}$\thinspace Mpc}

\newcommand{\vmh}{h^{-1}\mathrm{Mpc} }

\DeclareMathAlphabet{\pazocal}{OMS}{zplm}{m}{n}

\usepackage{amstext}

\begin{document}  

\title{
The Corona Borealis supercluster:\\
connectivity, collapse, and evolution \\ 
} 


\author {Maret~Einasto\inst{1} 
\and Rain~Kipper\inst{1} 
\and Peeter~Tenjes\inst{1} 
\and Heidi~Lietzen\inst{1}
\and Elmo~Tempel\inst{1} 
\and Lauri Juhan~Liivam\"agi\inst{1} 
\and Jaan~Einasto\inst{1,2,3}
\and Antti~Tamm\inst{1}
\and Pekka~Hein\"am\"aki\inst{4}
\and Pasi~Nurmi\inst{5}
}
\institute{Tartu Observatory, University of Tartu, Observatooriumi 1, 61602 T\~oravere, Estonia
\and
Estonian Academy of Sciences, Kohtu 6, 10130 Tallinn, Estonia
\and
ICRANet, Piazza della Repubblica 10, 65122 Pescara, Italy
\and 
Tuorla Observatory, Department of Physics and Astronomy, University of Turku, 20014
Turku, Finland
\and
Biodiversity Unit, University of Turku, 20014 Turku, Finland
}

\authorrunning{Einasto, M. et al. }

\offprints{Einasto, M.}

\date{ Received   / Accepted   }

\titlerunning{Corona Borealis}

\abstract
{
Rich superclusters of galaxies represent
dynamically active environments in which galaxies and their systems form and evolve. 
}
{We study the dynamical properties and  connectivity of the richest galaxy clusters in the Corona
Borealis (CB) supercluster and in the whole supercluster, and analyse 
star formation of galaxies in them
with the aim to understand the evolution of the supercluster and the
 galaxies within it. We compare it with the supercluster SCl~A2142.
}
{
We used the luminosity-density field to determine the high-density cores of the CB.
We identified the richest galaxy clusters in them and 
studied the dynamical state of the clusters, analysed their substructure,
 and studied the star formation properties of galaxies in them using normal mixture modelling and
the projected phase space diagram. 
We determined filaments in the supercluster to analyse the
connectivity of clusters. To understand the possible future evolution 
of the CB, we compared the mass distribution in it with predictions from the
spherical collapse model and analysed the gravitational acceleration field
in the CB.
}
{The richest clusters in the high-density cores of the CB are the Abell clusters
\object{A2065}, \object{A2061} (together with \object{A2067}), \object{A2089},
and Gr2064. 
At a radius $R_{\mathrm{30}}$  around each cluster (corresponding to the  density contrast
$\Delta\rho \approx 30$), the galaxy distribution shows a minimum.  
The  $R_{30}$ values for individual clusters lie in the range of $3 - 6$~\Mpc.
The radii of the clusters (splashback radii) lie
in the range of $R_{\mathrm{cl}} \approx 2 - 3$ ~$R_{\mathrm{vir}}$. 
The projected phase space diagrams and the comparison with the spherical collapse model 
suggest that $R_{\mathrm{30}}$ regions
have passed turnaround and are collapsing, forming infall regions around each cluster.
Galaxies in the richest cluster of the CB, A2065, and in its infall region 
have on average younger stellar
populations than other clusters and their environment.  
The cluster A2061 has the highest
fraction of galaxies with very old stellar populations, similar to those in A2142. The number of long filaments
that begin near clusters vary from one near A2089 to five near A2061.
The total connectivity of these clusters (the number of infalling groups and filaments)
varies from two to nine.
}
{
During the future evolution, the clusters in the main part of the CB may merge and form one of the largest
bound systems in the nearby Universe. Another part, with the cluster Gr2064, will form a separate system.
Our study suggests that structures with a current characteristic
density contrast $\Delta\rho \approx 30$ have passed turnaround and started to collapse 
at redshifts $z \approx 0.3 - 0.4$. 
The comparison of the number
and properties of the most massive collapsing supercluster cores from 
observations and simulations may serve as a test for cosmological models.
}


\maketitle

\section{Introduction} 
\label{sect:intro} 

The largest systems in the cosmic web that may eventually become gravitationally
bound  
are galaxy superclusters or their high-density cores, which are defined as the 
high-density regions in the cosmic web that
embed galaxies, galaxy groups, and
clusters. They are connected by filaments \citep{1953AJ.....58...30D, 1958Natur.182.1478D,
1978MNRAS.185..357J, 1980Natur.283...47E, 2000AJ....120..523R, 2011MNRAS.415..964L,
2015MNRAS.453..868O, 2015A&A...575L..14C}. 
Galaxy superclusters have been detected using various methods
such as the clustering analysis (e.g. the friends-of-friends method), the luminosity density field,
and the cosmological graph method
\citep{1993ApJ...407..470Z, 1994MNRAS.269..301E,
2001MNRAS.323...47B, 2003A&A...410..425E, 2003A&A...405..425E,
2004MNRAS.352..939E, 2011MNRAS.415..964L, 2014MNRAS.445.4073C, 2020MNRAS.493.5972H}. 
The study of superclusters, their dynamical state, and galaxy content
helps us to understand the formation and evolution of different structures
in the cosmic web, from individual galaxies to rich clusters and superclusters.

\citet{2014Natur.513...71T} used  data of the cosmic velocity field 
in the nearby Universe to show that galaxy flows from low-density regions around
the local (Laniakea) supercluster are converging. In this study 
the whole volume with converging galaxy flows (the basin
of attraction) and high-density region of galaxies is called the supercluster. 
\citet{2019A&A...623A..97E} proposed to call the low-density regions around traditional,
high-density regions in the cosmic web (superclusters) supercluster cocoons.
Superclusters act as great attractors. They
grow through the inflow of matter from surrounding cocoons.

Studies of superclusters have shown that rich superclusters have high-density
cores that may contain one or several rich clusters \citep{2007A&A...462..397E,
2007A&A...476..697E, 2015A&A...580A..69E, 2016A&A...595A..70E, 2020A&A...641A.172E}.
Rich clusters in supercluster cores may be merging or be in pre-merger state, for example, in the Shapley supercluster, in the Corona Borealis 
supercluster, or in the Sloan Great Wall superclusters 
\citep{1999ApJ...521...90H, 2005A&A...444..387D, 2005AdSpR..36..630B, 2013ApJ...779..189F,
2016A&A...595A..70E}. 
Merging clusters may also reside in poor superclusters
\citep{2001ApJ...562..254D, 2020arXiv201208491R, 1997A&AS..123..119E}.
Rich superclusters may have several cores with clusters
away from each other, as in the Perseus-Pisces supercluster or in  the Coma supercluster
\citep{1978MNRAS.185..357J, 2020A&A...634A..30M, 2020MNRAS.497..466S}.

The formation of the present-day structures in the cosmic web began with the growth of
tiny density perturbations in the very early Universe when smaller structures merged and accrete 
\citep{1978MNRAS.183..341W, 1980Natur.283...47E, 1996Natur.380..603B}. 
Rich galaxy clusters form where positive sections
of medium- and large-scale density perturbations combine
\citep{2011A&A...531A.149S}. 
Simulations show that present-day rich clusters 
have collected their  galaxies along filaments from regions with comoving radii 
of at least $10$~\Mpc\ at redshift $z \geq 1$ 
\citep{2013ApJ...779..127C, 2016A&ARv..24...14O}. 
The sizes of these regions around the clusters
depend on the cluster mass \citep{2013ApJ...779..127C} and also
on the assembly history of the infalling systems themselves. \citet{2013ApJ...779..127C} used these regions as one  criterion 
to search for protoclusters in the high-redshift Universe 
\citep[see also][]{2020MNRAS.496.3169A}. 
In the local Universe, \citet{2015A&A...580A..69E, 2020A&A...641A.172E} 
found a density minimum in the galaxy distribution
around the high-density core of the supercluster SCl~A2142.
\citet{2020A&A...641A.172E} found that within this region,
all galaxy systems (groups and substructures of A2142) are falling into the 
cluster \citep[see also][]{2018A&A...610A..82E, 2018A&A...620A.149E}.
They suggested that 
this region can be called the sphere of dynamical influence of the cluster,
and the density minimum marks the border of this 
sphere around  the main cluster of the supercluster in the supercluster high-density core.

Studies of superclusters and their high-density cores help us to understand the
coevolution of clusters and superclusters, and also of galaxies and groups in them. 
For example,
\citet{2003A&A...401..851E} and \citet{2012A&A...539A.106P} showed that poor groups of galaxies
near rich clusters are themselves richer and more luminous than poor groups
elsewhere. This environmental enhancement
of groups near clusters is evidence of the coevolution of groups
and clusters. The growth of clusters by infall of groups and galaxies
is accompanied by the changes in star formation properties of
galaxies \citep{2015ApJ...806..101H, 2017ApJ...844L..23C, 2018A&A...620A.149E,
2018MNRAS.473.2335M, 2018MNRAS.476.4877M, 2018ApJ...856...72O, 2019MNRAS.483.3227K, 2019A&A...621A.131M, 
2020ApJ...899...79S, 2020MNRAS.491.5406T}.
\citet{2020ApJ...888...89T} suggested that
the evolution of protoclusters and protosuperclusters may have two stages, 
one accompanied by the changes in star
formation dominated by the steady accretion of galaxies, and the other by
the merging between group-size halos, perhaps depending on the surrounding large-scale environments.

The studies of supercluster properties have revealed
that morphologically, superclusters can be divided into two classes:
spider-type and filament-type superclusters \citep{2007A&A...476..697E, 2011A&A...532A...5E}.
Spider-type superclusters have a complicated inner structure with galaxy clusters 
and groups connected by a large number of filaments. In contrast, in superclusters
with a filament morphology groups and clusters are connected by a small
number of filaments.
\citet{2014A&A...562A..87E} and \citet{2017ApJ...835...56C} showed that
spider-type superclusters have a higher fraction of star-forming galaxies
and clusters with more substructure than filament-type
superclusters. This is again evidence of a coevolution of superclusters and
galaxy groups and galaxies within them.

In this paper we focus on the study of the  rich nearby Corona Borealis (CB) supercluster
at redshift $z \approx 0.07$. 
The richest clusters in the CB supercluster are \object{A2065}, \object{A2061},
\object{A2067}, and \object{A2089}.
The CB supercluster is located at
the crossing of three chains of rich superclusters, in the 
dominant supercluster plane 
\citep{1997A&AS..123..119E, 2011A&A...532A...5E}. Morphologically,
the CB supercluster is elongated and a clumpy
supercluster of the multispider type \citep{2011A&A...532A...5E}.

The dynamics of the CB was first studied by \citet{1988AJ.....95..267P}
who concluded that mass of the supercluster  is probably
sufficient to bind the system.
\citet{1998ApJ...492...45S} used data from the Norris Survey  to show that the most 
prominent core region of 
the CB supercluster centred at A2065 may have started to collapse.
\citet{2014MNRAS.441.1601P} suggested that if the intercluster mass in the CB were sufficient,
then it might be collapsing and forming an extended bound structure.
The CB supercluster is listed among the future virialised structures
by \citet{2011MNRAS.415..964L}. 

The interest in the study of the CB supercluster increased when it was found that
the Cosmic Microwave Background (CMB) cold spots lie in the direction of the CB 
\citep{2008MNRAS.391.1127G, 2009MNRAS.396...53P,
2010MNRAS.403.1531G, 2010hsa5.conf..329P}. 
These studies found that the CB cold spot
may be partly related to the warm-hot diffuse gas in the supercluster
region between galaxy filaments, but it may also be related to some
distant cluster or some other large structures.

The aim of this study is to understand the evolution of the CB supercluster and its components,
the connections between clusters in the
supercluster, and transformations of galaxies in it.
We focus on the study of the high-density cores of the
supercluster, which may be local centres of attraction. These centres evolve by infall 
of galaxies and galaxy groups and clusters. If the supercluster contains several high-density cores, then each of them may collapse at present or in the future. This means
that the supercluster may be split into several systems in the future. It is also
possible that cores will join to form massive bound object(s) in the future.
Therefore one goal of our study is to distinguish between these future
scenarios for the CB supercluster.
We also wish to understand whether dynamical processes in clusters and in the supercluster
scale are accompanied with the change in star formation properties of
the galaxies in them.

We determine the high-density cores of
the CB using the luminosity-density field  based on galaxy data from the Sloan Digital Sky Survey 
(SDSS) and  select the richest clusters from the high-density cores of the supercluster
as possible centres of attraction. 
Next we analyse the distribution of galaxies around them to detect their
regions of dynamical influence. 
We analyse the cluster substructure with normal mixture modelling and use 
the Projected Phase Space (PPS) diagrams to study the possible infall of galaxies, groups, and
filaments into clusters.
We use data of filaments to study the connectivity of clusters.  
We also study galaxy populations in clusters and between them,
with an emphasis on the star formation properties
of galaxies and on the galaxies in various stages of transformation, that is, red star-forming
galaxies and recently quenched galaxies. Finally, to predict the 
evolution of clusters and the whole supercluster,
we compare the mass distribution
around them with the predictions from the spherical collapse model
and analyse the acceleration field of the supercluster. 
A detailed study of galaxy populations in 
groups of the CB will be presented separately.

One aim of this paper is to compare the properties of the richest clusters in the CB
with the properties of cluster \object{A2142}, which is the richest cluster in the supercluster
SCl~A2142 \citep{2015A&A...580A..69E, 2020A&A...641A.172E}. 
\citet{2016ApJ...827L...5M} found that cluster A2142 has a very high number of luminous galaxies
that are not reproduced in simulations. In addition, 
\citet{2018A&A...610A..82E} showed that cluster A2142 is an outlier in the mass - richness
diagram for groups and clusters from the
\citet{2014A&A...566A...1T} group catalogue at nearly  the same distances.
A2142 has a significantly higher fraction
of red, passive galaxies than other rich clusters at the same distances.
In this study we compare the galaxy content and dynamical properties, and
the connectivity of A2142 and the richest clusters in the CB. A comparison of individual
clusters may reveal details that are lost when an ensemble of clusters is compared,
as in \citet{2018A&A...610A..82E}. 

We use the following cosmological parameters: the Hubble parameter $H_0=100~ 
h$ km~s$^{-1}$ Mpc$^{-1}$, the matter density $\Omega_{\rm m} = 0.27$, and 
the dark energy density $\Omega_{\Lambda} = 0.73$ 
\citep{2011ApJS..192...18K}.

\section{Data} 
\label{sect:data} 

\subsection{Supercluster, group, and filament data}
\label{sect:gr}

\begin{figure}
\centering
\resizebox{0.45\textwidth}{!}{\includegraphics[angle=0]{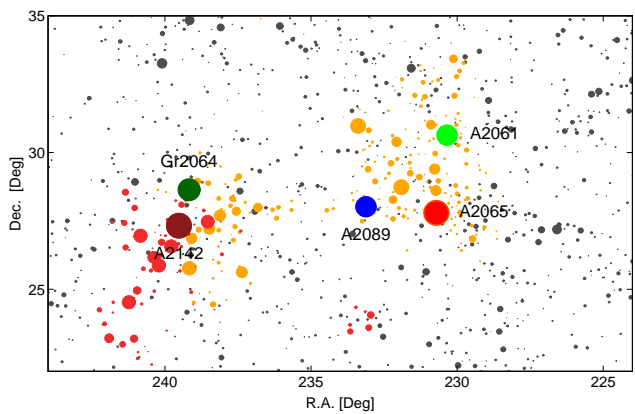}}
\caption{
Distribution of galaxy groups in a large region around the CB in the plane of the sky.   
The size of the circles is proportional to group size in the sky.
Grey circles show the positions of groups in a low global density environment,
$D8 < 5$. Orange circles correspond to groups in the CB supercluster, and 
red circles show groups in a high global density environment with $D8 \geq 8$
that do not belong to the CB.
These groups mostly lie in supercluster SCl~A2142. 
In the sky projection, part of the CB supercluster is projected
on SCl~A2142, and some groups from these superclusters overlap in the figure.
Large coloured circles indicate the
location of the richest clusters with corresponding ID numbers.
}
\label{fig:cbwideradec}
\end{figure}

We used data from the supercluster, group,  and filament catalogues by 
\citet{2012A&A...539A..80L} and \citet{2012A&A...540A.106T, 2014A&A...566A...1T,
2014MNRAS.438.3465T, 2016A&C....16...17T}. 
The initial data of galaxies with which these catalogues were generated were 
taken from the SDSS DR10 MAIN spectroscopic galaxy sample  with 
apparent Galactic extinction-corrected $r$ magnitudes $r \leq 
17.77$ and redshifts $0.009 \leq z \leq 0.200$
\citep{2011ApJS..193...29A, 2014ApJS..211...17A}.
The catalogues of galaxy superclusters, groups, and filaments are
available from the database of cosmology-related catalogues 
at \url{http://cosmodb.to.ee/}.
Data of the galaxy properties were taken from the SDSS DR10 web page  
\footnote{\url{http://skyserver.sdss3.org/dr10/en/help/browser/browser.aspx}}.
The same data were used by \citet{2018A&A...610A..82E, 2020A&A...641A.172E} in the analysis
of supercluster SCl~A2142.

\begin{table*}[ht]
\caption{General properties of the CB supercluster.}
\label{tab:cbdata}  
\begin{tabular}{rrrrrrrrrr} 
\hline\hline 
\multicolumn{1}{c}{(1)}&(2)&(3)&(4)& (5)&(6)&(7)& (8)&(9)\\      
\hline 
\multicolumn{1}{c}{ID}& $N_{\mbox{gal}}$ & $d_{\mbox{peak}}$ &$M$  
 & $\mathrm{Diam}$ & $D8_{\mathrm{max}}$ & 
$N^{\mathrm{gr}}_{\mathrm{4}}$ & $N^{\mathrm{gr}}_{\mathrm{2-3}}$ & $N_{\mathrm{1}}$ \\
& &  [$h^{-1}$ Mpc] & [$h^{-1}M_\odot$]&[$h^{-1}$ Mpc] & \\
\hline
 230+027+007 & 2047  & 215 &$4.3\times~10^{15}$ &  55 & 11.5 & 88 & 103 & 296 \\ 
\hline
\end{tabular}\\
\tablefoot{                                                                                 
Columns in the Table are as follows:
(1): supercluster ID AAA+BBB+ZZZ, where AAA is R.A., +/-BBB is Dec., and ZZZ is 100$z$;
(2): the number of galaxies in the CB, $N_{\mbox{gal}}$;
(3): the distance of the density maximum, $d_{\mbox{peak}}$;
(4): the mass of the CB, $M$;
(5): the supercluster diameter (the maximum distance between galaxies in
the CB), $\mathrm{Diam}$;
(6): the maximum value of the luminosity-density field in the CB, calculated with
the $8$~\Mpc\ smoothing kernel, $D8_{\mathrm{max}}$, in units of the mean density as described in the text.
(7): Number of groups with $\geq 4$ member galaxies, $N^{\mathrm{gr}}_{\mathrm{4}}$;
(8): Number of groups with $2 - 3$ member galaxies, $N^{\mathrm{gr}}_{\mathrm{2-3}}$;
(9): number of single galaxies, $N_{\mathrm{1}}$;
}
\end{table*}

\begin{figure}
\centering
\resizebox{0.45\textwidth}{!}{\includegraphics[angle=0]{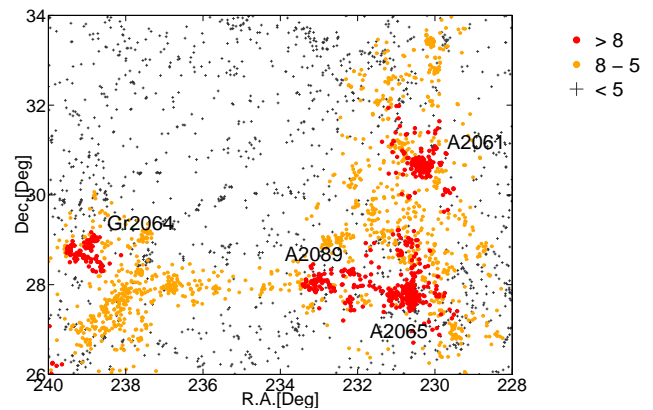}}
\caption{
Distribution of galaxies in the CB supercluster region in the plane of the sky.   
The colour marks the global luminosity density $D8$. 
Red circles correspond to the galaxies in high-density cores with $D8 \geq 8$,
orange circles  show galaxies in the outskirts of the supercluster with $5 \leq D8 < 8$,
and grey symbols show galaxies in the low-density region around the supercluster
with $D8 < 5$. The ID numbers of the richest
clusters in the high-density cores of the supercluster 
are shown (Abell clusters A2065, A2061, and A2089, and cluster Gr2064 that does not belong to Abell).
}
\label{fig:cbradec}
\end{figure}

In the catalogue of galaxy superclusters by \citet{2012A&A...539A..80L},
superclusters are determined on the basis of the luminosity-density  field.
In this study, the luminosity-density field was calculated using a
$B_3$ spline kernel with a smoothing length 8~\Mpc\
at a location $x$,
\begin{equation}
    B_3(x) = \frac{1}{12} \left(|x-2|^3 - 4|x-1|^3 + 6|x|^3 - 4|x+1|^3 + |x+2|^3\right).
\end{equation}
Superclusters were defined as the connected 
volumes above a threshold density $D8 = 5.0$
(in units of mean 
luminosity density, $\ell_{\mathrm{mean}}$ = 
1.65$\cdot10^{-2}$ $\frac{10^{10} h^{-2} L_\odot}{(\vmh)^3}$
in the luminosity density field). 
The threshold density for determining superclusters was chosen as a result of the
analysis of supercluster properties when the threshold density level was varied.
This analysis showed that at high threshold densities, the method selects only the highest density regions as superclusters, which can be considered high-density cores of superclusters. If the threshold density level is low,
then neighbouring superclusters are combined into huge complexes of superclusters,
as was shown in the case of the Sloan Great Wall in \citet{2011ApJ...736...51E}.
For details of the luminosity density field and 
the supercluster definition, we refer to \citet{2012A&A...539A..80L}.

The data for the CB supercluster as defined 
in \citet{2012A&A...539A..80L} are given in Table~\ref{tab:cbdata}. 
The length of the supercluster
is $\approx 55$~\Mpc, and its mass is $M \approx 4.3\times~10^{15}h^{-1}M_\odot$.
In Fig.~\ref{fig:cbwideradec} we show the sky distribution of galaxy groups in the CB and
its surrounding region, which also includes the supercluster SCl~A2142. SCl~A2142
is partly projected on the Gr2064 part of CB, and is 
connected with it by a filament through 
a void between superclusters
\citep{2007AstL...33..211K, 2019NewA...69....1P, 2020A&A...641A.172E}.

We used data for {\it \textup{flux-limited galaxy groups}} from the catalogue 
by \citet{2014A&A...566A...1T} to find groups and clusters that belong 
to the CB. \citet{2014A&A...566A...1T} 
determined galaxy groups using 
the friends-of-friends cluster analysis 
method \citep{1982Natur.300..407Z, 1982ApJ...257..423H}. In this method,
a galaxy is considered a member of a group 
if at least one group member galaxy lies closer to this galaxy 
than a linking length. In a flux-limited sample, the density of galaxies slowly 
decreases with distance. To take this selection effect properly into account 
when a group catalogue was constructed, the 
linking length was rescaled with distance, calibrating the scaling relation by observed 
groups. As a result, the 
maximum sizes in the sky projection and the velocity dispersions of the groups 
are similar at all distances. The  redshift-space distortions (also known as Fingers of God) 
for groups were supressed,
as described in detail in \citet{2014A&A...566A...1T}.
\citet{2014A&A...566A...1T} reported details of the data 
reduction, the group finding procedure, and the description of the group catalogue.
We determined the number of groups as at least four member galaxies and with $  \text{two to three}$
member galaxies in Table~\ref{tab:cbdata}. In Table~\ref{tab:cbdata} we also list the number
of single galaxies, that is, galaxies that do not
have neighbours in the galaxy catalogue that are luminous enough to be included in
the SDSS spectroscopic sample.
In the flux-limited group catalogue, approximately 52\% of all galaxies 
are single \citep{2009A&A...495...37T}. However, single galaxies are more
common in the low-density environment outside superclusters. In
superclusters, the fraction of single galaxies is lower, and in the CB, approximately
15\% of all galaxies do not belong to any group. \citet{2009A&A...495...37T} also showed based on the luminosity functions of
galaxies that especially in the supercluster
environment, single galaxies may be the brightest galaxies of faint groups,
and truly isolated galaxies may lie in low-density regions, but these are rare in superclusters.
We note that one fossil group candidate
is included in the groups in the CB, at the edge of the supercluster
at approximately $R.A. =232.2$\degr and $Dec. = 32.4$\degr,
with a magnitude gap between the brightest and second brightest galaxy 
$|\Delta M_{12}| = 2.2$. Groups with such large magnitude gaps are defined
as possible fossil groups in X-ray studies  \citep{1994Natur.369..462P}.

As candidates for the centres of attraction in the high-density cores of
the supercluster, we chose the richest clusters above the global luminosity density level
$D8 \geq 8$. 
This threshold density level for high-density cores of CB was chosen
on the basis of earlier studies of superclusters, which showed that rich superclusters
contain high-density cores, while such cores are absent in poor superclusters
\citep{2007A&A...464..815E, 2007A&A...462..397E, 2015A&A...580A..69E,
2020A&A...641A.172E}.
We searched for rich clusters from the group catalogue 
with $D8 \geq 8$, and selected the richest of them
as possible centres of attraction in the supercluster.
This choice returned Abell clusters \object{A2065}, 
\object{A2061}, 
and \object{A2089}, and one rich cluster without an Abell ID, Gr2064 in the \citet{2014A&A...566A...1T}
group catalogue.
The data of the rich clusters are given in Table~\ref{tab:cl}.
In this table we also list the data of the low-mass cluster A2067,
which, as we show in Sect.~\ref{sect:a2061}, forms a pair of
merging clusters with A2061.  
The sky distribution of galaxies in the CB supercluster is shown in
Fig.~\ref{fig:cbradec}. In this figure galaxies are colour-coded according to the
luminosity-density $D8$ value around them.
The high-density cores have a threshold density $D8 \geq 8$,
and the outskirts region of the supercluster is defined as having
$5 \leq D8 < 8$.
Figure~\ref{fig:cbradec} shows that the CB supercluster consists of two parts,
divided at  $R.A. \approx 235$\degr. The CB is commonly associated with
the main part with $R.A. < 235$\degr. Another part of the 
CB is connected with the main part by a bridge
of galaxies, and there are no Abell clusters in this part. The richest cluster there is Gr2064.
In the following, we call these parts the main part and the Gr2064 part of the CB.

We also identified galaxy filaments in the CB. 
\citet{2014MNRAS.438.3465T, 2016A&C....16...17T} detected filaments by 
applying a marked point process to the galaxy distribution (the Bisous model).
For each galaxy, a distance from the nearest filament
axis was calculated.  
Galaxies were considered filament members 
when their distance from the nearest filament axis was within $0.5$~\Mpc\
\citep[see][for details]{2014MNRAS.438.3465T, 2020A&A...641A.172E}.

\begin{table*}[ht]
\caption{Data of rich galaxy clusters in the CB.}
\begin{tabular}{rrrrrrrrrrrrr} 
\hline\hline  
(1)&(2)&(3)&(4)&(5)& (6)&(7)&(8)&(9)&(10)&(11)&(12)&(13)\\      
\hline 
No. & Abell ID& ID &$N_{\mathrm{gal}}$& $\mathrm{R.A.}$ & $Dec.$ 
&$\mathrm{Dist.}$ &$\sigma_{v}$ &  $R_{\mathrm{vir}}$&  $R_{\mathrm{max}}$ & $M_{\mathrm{dyn}}$  
& $L_{\mathrm{tot}}$ & $D8$ \\
\hline                                                    
 1 &2065& 7045 & 161& 230.7 &27.7&  213 & 1082&  0.7& 3.3 &1.53  &1.9 & 10.4 \\
 2 &2061& 2263 & 107& 230.3 &30.6&  230 &  690&  0.5& 1.4 &0.39  &1.6 & 10.3 \\
 3 &2089& 3278 &  67& 233.1 &28.0&  218 &  450&  0.6& 1.8 &0.22  &0.9 &  8.4 \\
 4 &    & 2064 &  77& 239.1 &28.6&  229 &  447&  0.7& 2.4 &0.34  &1.3 &  8.3 \\
\hline                                        
 5 &2067& 2109 &  41& 230.9 &31.0&  219 &  383&  0.5& 1.2 &0.14  &0.7 &  5.6    \\
\hline                                        
\label{tab:cl}  
\end{tabular}\\
\tablefoot{                                                                                 
Columns are as follows:
(1): Order number of the cluster;
(2): Abell ID number of the cluster;
(3): ID of the cluster from \citet{2014A&A...566A...1T};
(4): Number of galaxies in the cluster, $N_{\mathrm{gal}}$;
(5)--(6): cluster centre right ascension and declination (in degrees);
(7): cluster centre comoving distance (in $h^{-1}$ Mpc);
(8): line-of-sight velocity dispersion of cluster member galaxies;
(9): cluster virial radius (in $h^{-1}$ Mpc);
(10): maximum radius of a cluster in the plane of the sky (in $h^{-1}$ Mpc);
(11): dynamical mass of the cluster assuming an NFW density profile, $M_{\mathrm{dyn}}$,
(in $10^{15}h^{-1}M_\odot$);
(12): cluster total luminosity (in $10^{12} h^{-2} L_{\sun}$);
(13): luminosity-density field value at the location of the cluster, $D8$, 
in units of the mean density as described in the text.
}
\end{table*}

\subsection{Galaxy populations}
\label{sect:galpop}

We used data of the galaxy properties 
from the SDSS DR10 web page\footnote{\url{http://skyserver.sdss3.org/dr10/en/help/browser/browser.aspx}}.
From the available data for each galaxy, we used the magnitudes, colours,
stellar masses $M^{\mathrm{*}}$,  star formation rates (SFR),
and the $D_n(4000)$ index.

The absolute magnitudes of the galaxies were calculated as
\begin{equation}
M_r = m_r - 25 -5\log_{10}(d_L)-K,
\end{equation} 
where $d_L$ is the luminosity distance in units of $h^{-1}$Mpc and
$K$ is the $k$+$e$-correction, calculated as in 
\citet{2007AJ....133..734B} and  \citet{2003ApJ...592..819B}
\citep[see][for details]{2014A&A...566A...1T}.
To calculate the galaxy rest frame colour  $(g - r)_0$ and the concentration 
index, we used galaxy magnitudes from the SDSS photometric data. 
Red and blue galaxies are
approximately separated by the colour index value $(g - r)_0 = 0.7$; red galaxies 
have $(g - r)_0 \geq 0.7$ \citep{2014A&A...562A..87E}.

Data of stellar masses $M^{\mathrm{*}}$,  SFR,
and the $D_n(4000)$ index for galaxies were taken from 
the Max Planck for Astrophysics (MPA) and Johns Hopkins University (JHU) 
spectroscopic catalogue \citep{2004ApJ...613..898T, 2004MNRAS.351.1151B}.  
In this catalogue the properties of 
galaxies are calculated using 
the stellar population synthesis models and fitting SDSS photometry and spectra 
with \citet{2003MNRAS.344.1000B} models.
The stellar masses of galaxies are calculated 
as described in \citet{2003MNRAS.341...33K}.

\begin{figure}[ht]
\centering
\resizebox{0.44\textwidth}{!}{\includegraphics[angle=0]{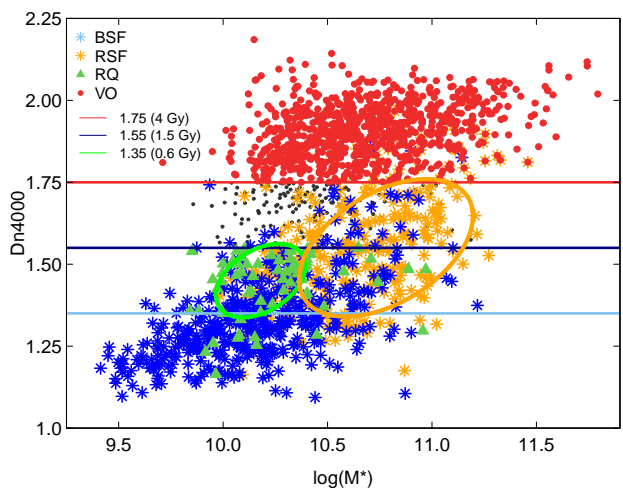}}
\caption{
$D_n(4000)$ index vs. stellar mass $\log M^{\mathrm{*}}$ for galaxies in the CB
supercluster.
Blue stars denote BSF galaxies,
orange stars denote RSF galaxies, green triangles mark
RQ galaxies, black dots show red galaxies with $1.55 \leq D_n(4000) < 1.75$,
and red filled circles 
denote VO galaxies (see Table\ref{tab:galpopdef}). 
The green ellipse approximately borders the area populated by RQ galaxies,
and the orange ellipse shows the area populated by RSF galaxies.
}
\label{fig:cbd4smass}
\end{figure}

$\text{The }D_n(4000)$ index is the ratio of the average flux densities
in the band $4000 - 4100 \angstrom$ and $3850 - 3950 \angstrom$.
It is correlated with the time that has passed 
from the most recent star formation event in a galaxy. The
$D_n(4000)$ index can be used as a proxy for the age of stellar populations of galaxies
and star formation rates.
We used the $D_n(4000)$ index of galaxies as calculated in \citet{1999ApJ...527...54B}.
\citet{2003MNRAS.341...33K} showed that the value $D_n(4000) = 1.55$ corresponds to 
a mean age of about $1.5$~Gyr.
This limit can be used to separate galaxies with old and young
stellar populations \citep[see][]{2003MNRAS.341...54K, 2017A&A...605A...4H, 2018A&A...620A.149E}.
Galaxies with young stellar populations have  $D_n(4000) \leq 1.55$. 
According to \citet{2003MNRAS.341...33K}, the value $D_n(4000) = 1.75$ corresponds to 
a mean age of about $4$~Gyr (for solar metallicity) or older 
(for lower metallicities). 
At $D_n(4000) = 1.75,$  the galactic star formation rate drops 
\citep[see Fig. 27 in ][]{2004MNRAS.351.1151B}. 
The value $D_n(4000) = 1.35$ limits galaxies with very young stellar populations
with a mean age of approximately  $0.6$~Gyr only \citep{2003MNRAS.341...33K}.

In our study we divided galaxies into various populations according to their
star formation properties. 
Galaxies with very old stellar populations (hereafter VO galaxies) are defined as those
with $D_n(4000) \geq 1.75$.
Blue star forming (hereafter BSF) galaxies,
red star forming (RSF), and recently quenched (RQ) galaxies are defined 
using parameter combinations.
BSF galaxies are defined as galaxies with blue colours 
($(g - r)_0 < 0.7$)  and a high star formation rate ($\log \mathrm{SFR} \geq -0.5$). 
RSF galaxies are defined  as galaxies with colour index
$(g - r)_0 \geq 0.7$  and a star formation rate $\log \mathrm{SFR} \geq -0.5$.
RQ galaxies are defined as galaxies with low SFRs that have young stellar populations
(low values of the $D_n(4000)$ index,  
$D_n(4000) \leq 1.55$ and  $\log \mathrm{SFR} < -0.5$). 
The data of the galaxy populations are summarised in Table~\ref{tab:galpopdef}.

Figure~\ref{fig:cbd4smass} shows galaxies from different populations
in a $D_n(4000)$ index versus stellar mass $\log M^{\mathrm{*}}$ diagram.
Galaxies in transformation
(RSF and RQ)  form a sequence in this plot in which galaxies with higher stellar 
mass are red but are still forming their stars, while galaxies with lower
stellar mass from the same $D_n(4000)$ index range 
are already quenched. As discussed in \citet{2018A&A...620A.149E} and in
Einasto et al. (2020), the location of RSF and RQ galaxies in the cosmic web is somewhat different,
RSF galaxies follow
the distribution of groups and inner parts of filaments more closely.
RQ galaxies lie mostly in the infall regions of groups or in the outskirts of filaments. 
Galaxies below $D_n(4000) = 1.35$ line are almost all BSF galaxies with low stellar mass
and very young stellar populations.
Galaxies from the stellar mass range 
$M^{\mathrm{*}} \approx 10^{10}h^{-1}M_\odot -10^{11}h^{-1}M_\odot$
may belong to any of these populations (Fig.~\ref{fig:cbd4smass}).
In what follows we examine where galaxies from different populations
lie in the CB supercluster, and where galaxy transformations occur.

\begin{table*}[ht]
\caption{Data of galaxy populations.}
\begin{tabular}{lllr} 
\hline\hline  
(1)&(2)&(3)&(4)\\      
\hline 
Population & Abbr. &Definition&  $N_{\mathrm{gal}}$ \\
\hline                                                    
Blue star-forming galaxies  & BSF &  $(g - r)_0 < 0.7$, $\log \mathrm{SFR} \geq -0.5$      & 438 \\
Red star-forming galaxies   & RSF &  $(g - r)_0 \geq 0.7$, $\log \mathrm{SFR} \geq -0.5$   & 303 \\
Recently quenched galaxies  & RQ  &  $D_n(4000) \leq 1.55$,  $\log \mathrm{SFR} < -0.5$    &  72\\
Galaxies with very old stellar populations & VO &  $D_n(4000) \geq 1.75$ & 892 \\
\hline
\label{tab:galpopdef}  
\end{tabular}\\
\tablefoot{                                                                                 
Columns are as follows:
(1): Galaxy population;
(2): abbreviation;
(3): definition of a given population;
(4): number of galaxies in a given population in the CB. 
}
\end{table*}

\section{Methods} 
\label{sect:met}

\subsection{Spherical collapse model} 
\label{sect:sph} 

The spherical collapse model describes the evolution of a spherical 
perturbation in an expanding universe \citep{1980lssu.book.....P,
1984ApJ...284..439P, 1991MNRAS.251..128L}.  
In standard models with a cosmological constant, the 
acceleration of the expansion of the Universe 
began at the redshift 
$z \approx 0.5$ by the influence of dark energy \citep{2008ARA&A..46..385F}.   
The formation of structure slowed down, and   
the largest bound structures just began forming.  
The evolution of a collapsing spherical shell is determined by the mass
in its interior, and it can be characterised by several important
epochs \citep{2015A&A...575L..14C, 2015A&A...581A.135G}.

For a spherical volume, 
the density ratio to the mean density (overdensity)
$\Delta\rho = \rho/\rho_{\mathrm{m}}$ can be calculated as
\begin{equation}
\Delta\rho=6.88\,\Omega_\mathrm{m}^{-1}\left(\frac{M}{10^{15}h^{-1}M_\odot}\right)
        \left(\frac{R}{5h^{-1}\mathrm{Mpc}}\right)^{-3}.
\label{eq:sph}
\end{equation}
From Eq.~(\ref{eq:sph}), we determine the mass of a structure as
\begin{equation}
M(R)=1.45\cdot10^{14}\,\Omega_\mathrm{m}\Delta\rho\left(R/5h^{-1}\mathrm{Mpc}\right)^3h^{-1}M_\odot.
\label{eq:mass1}
\end{equation}

The spherical collapse model defines several epochs
in the evolution of a perturbation as follows. 
Turnaround (T) corresponds to the epoch 
at which a spherical overdensity region
decouples from expansion and its collapse begins,
with  $\Delta\rho = 13.1$ \citep[for the cosmological
parameters used in this paper, see e.g. ][]{2015A&A...581A.135G}. 
Overdensity regions with $\Delta\rho = 8.73$
will collapse in the future (FC for future collapse).
The density contrast $\Delta\rho = 5.41$ corresponds to so-called zero 
gravity (ZG), at which the radial peculiar velocity component of 
the test particle velocity equals the Hubble expansion,
and gravitational attraction of the system and its expansion are equal.
The density contrast $\Delta\rho =  1$ 
corresponds to a linear mass scale or the Einstein-Straus radius at which
the radial velocity around a system reaches the Hubble velocity,
$u = HR,$ and peculiar velocities $v_{\mathrm{pec}} = 0$
\citep{2015A&A...577A.144T, 2015A&A...581A.135G}. Around superclusters, the 
linear mass scale approximately follows the cocoon borders \citep{2020A&A...641A.172E}.

\subsection{Mass distribution in the CB supercluster} 
\label{sect:mass} 

In our study we compared the density contrast around the richest galaxy clusters
in the CB supercluster with the predictions
of the spherical collapse model to understand the possible evolution of clusters and the whole 
supercluster. 
To calculate the density contrast, we calculated the distribution of mass around 
the rich clusters in the supercluster.
The supercluster itself was determined as the overdensity region in the luminosity-density
field, as described in Sect.~\ref{sect:gr}. However, the analysis of the mass-to-light ratios
of galaxy systems shows that the luminosity-density field is a biased
tracer of the underlying mass field \citep{2014MNRAS.439.2505B, 2016A&A...595A..70E}. Thus
we cannot use the luminosity-density
field directly to determine the mass distribution in the supercluster. 

To determine the mass distribution in the supercluster, we therefore used
the dynamical mass of galaxy groups from \citet{2014A&A...566A...1T}.
In the case of groups with at least four member galaxies, we directly used the masses from the
\citet{2014A&A...566A...1T} group catalogue.
For very poor groups with fewer than four member galaxies,
the mass estimates from \citet{2014A&A...566A...1T} have a large scatter, which makes the
masses of individual groups  unreliable. Therefore we used the 
median mass of groups with $  \text{two to three}$ member galaxies to calculate the total mass of groups
within a given clustercentric distance,  
$M \approx 1.6\times~10^{12}h^{-1}M_\odot$. 
The median mass was multiplied by the number of groups of this richness class
to obtain the total mass in these groups at a given clustercentric distance interval around
clusters.
To estimate  the mass related to single galaxies, we considered single galaxies
as the brightest galaxies of faint groups with other group members being
too faint to be included in the SDSS spectroscopic sample. To take the mass of these 
faint groups into account, we used the median mass of very poor groups with $  \text{two to three}$ member galaxies.  
We could also have applied the stellar mass - halo mass relation to determine the 
mass of dark matter haloes around single galaxies, as was done for example in
\citet{2016A&A...595A..70E} and in \citet{2018A&A...610A..82E}, using 
the stellar mass - halo mass relation from \citet{2010ApJ...710..903M}, and drop the
assumption about faint groups. However, as shown in
\citet{2016A&A...595A..70E, 2018A&A...610A..82E}, these two estimates agree well on average (within adopted mass errors), therefore we used the median mass of poor groups.
The details of this method were described in \citet{2015A&A...580A..69E} and
in \citet{2018A&A...620A.149E}, who applied the method to calculate the mass of supercluster 
SCl~A2142. The mass distribution around each cluster was calculated by summing the masses of clusters and groups
within a sphere of increasing radius. As in \citet{2018A&A...620A.149E}, we used a 50\% mass error.
Superclusters also contain gas. The gas mass may form approximately 10\% 
of the supercluster mass \citep{2016A&A...592A...6P}. 
To obtain the total mass of the supercluster, we added this to the
sum of the dynamical masses of groups. Rich groups with at least four member galaxies 
give approximately 75\% of the total mass of the supercluster,
and poor groups and single galaxies 10\% and 5\%, respectively.

The accuracy of the supercluster mass determination can be tested with simulations.
For group dynamical masses comparison of group masses from the
\citet{2014A&A...566A...1T} group catalogue and from the mock group catalogues
from simulations showed that there is no clear bias between these
masses \citep{2014MNRAS.441.1513O, 2015MNRAS.449.1897O}.
The mass of a supercluster calculated using the dynamical masses
of groups and the gas mass is biased relative to the
total mass of the supercluster. This bias depends on the richness limit 
of the groups that are used in the calculations \citep{2014A&A...567A.144C}.
\citet{2015A&A...580A..69E} compared supercluster mass estimates 
for the A2142 supercluster and for simulated superclusters \citep{2014A&A...567A.144C},
and estimated that the bias factor may be of order of $1.8$, which means
that we are still underestimating the mass of the supercluster.
If we only used rich groups with reliable mass estimations, the
bias factor would be larger. 
We assume that the missing mass partly comes from unobserved faint
galaxies and groups, which can be uncovered in simulated superclusters
\citep{2014A&A...567A.144C}. If approximately 50\% of all galaxies in the supercluster
were
faint single galaxies, then the total mass of the supercluser would increase
by approximately 10\%, which is within the adopted mass errors.

\subsection{Substructure of clusters} 
\label{sect:sub} 

We applied multidimensional normal mixture modelling 
to search for the 3D substructure of galaxy clusters.
We also analysed whether the structures of clusters and 
their galaxy populations are related with (possibly infalling) 
groups and filaments near clusters.  
In this analysis we used the package {\it mclust} for classification and clustering
\citep{fraley2006} from {\it R}  statistical environment 
\citep[][\texttt{http://www.r-project.org}]{ig96}.
This package studies a  finite mixture of distributions, 
in which each component is taken to correspond to a 
different subgroup of the cluster.
With {\it mclust} we searched for an optimal model for the clustering of the data
among the models with varying shape, orientation, and volume, 
and determined the optimal number of  components in the data and the membership
of components (classification of the data). 
{\it mclust} also calculates the uncertainty of the classification, which is defined as one minus the 
highest probability of a datapoint to belong to a component. 
It finds for each datapoint 
the probability to belong to a component. 
The mean uncertainty
for the full sample  is a statistical estimate of the reliability
of the results.
As input for {\it mclust,} we used 
the sky coordinates and line-of-sight velocities 
(calculated from their redshifts) of the cluster member galaxies. 
The velocity values were scaled to make them of the same order as the values
of coordinates.  The best solution for the components 
was chosen using the Bayesian information criterion (BIC). 

\subsection{Projected phase space diagram} 
\label{sect:pps} 

To analyse the galaxy content and merging history of clusters, we employed 
the PPS. The PPS diagram shows line-of-sight velocities of galaxies 
with respect  to the cluster mean velocity versus projected clustercentric distance.
Simulations show that in the PPS diagram, galaxies at small clustercentric 
radii form an early infall (virialised) population with 
infall times $\tau_{\mathrm{inf}} > 1$~Gyr, and galaxies 
at large clustercentric radii 
form late or ongoing infall populations with $\tau_{\mathrm{inf}} < 1$~Gyr
\citep{2013MNRAS.431.2307O, 2014ApJ...796...65M, 2015ApJ...806..101H, 
2017ApJ...843..128R, 2019MNRAS.484.1702P, 2019ApJ...876..145S}. 
Early and late infall regions can be approximately 
separated with a line as follows \citep{2013MNRAS.431.2307O}:
\begin{equation}\label{eq:infalltime}
\left\lvert \frac{v-v_\text{mean}}{\sigma_\text{cl}} \right\lvert= 
-\frac{4}{3}\frac{D_\text{c}}{R_\text{vir}}+2 , 
\end{equation}
where $v$ are the velocities of the galaxies, $\sigma_\text{cl}$ is the velocity dispersion of galaxies in
the cluster,
$D_\text{c}$ is the projected clustercentric distance, and 
$R_{\mathrm{vir}}$ is the cluster virial radius. 
\citet{2013MNRAS.431.2307O} showed that in simulations a large percentage of 
galaxies  on the left of the line at small clustercentric 
radii form an early infall (virialised) population 
with $\tau_{\mathrm{inf}} > 1$~Gyr.
However, simulations show that  galaxies with late infall time 
may populate areas of PPS diagram at all projected radii and 
velocities, which complicates the interpretation
of the PPS diagram from observations \citep{2015ApJ...806..101H}. 

In our study we used PPS diagrams to study the 
dynamical state of clusters, the distribution
of galaxies with different star formation properties in them, and
to detect possible infalling galaxies, groups, and filaments.

\subsection{Cluster radii $R_{\mathrm{cl}}$.} 
\label{sect:rcl} 

In order to help us to interpret the PPS diagrams, we additionally used 
the radius that we denote $R_{\mathrm{cl}}$. This radius corresponds to the radius
of a cluster for one-component clusters and to the radius of the main component
in clusters with several components \citep[for cluster A2142, see][]{2018A&A...610A..82E}. 
At this radius, the outer (infalling) components of the cluster enter the main cluster,
and therefore this radius defines the infall zone at the cluster boundaries,
which are part of the late infall region in the PPS diagram.
The infall zone of the cluster is also the zone in which splashback galaxies are to be most likely found,
that is, galaxies that entered the cluster a long time ago and  might now be moving
out of the cluster. The radius that we call  $R_{\mathrm{cl}}$
is commonly known as the splashback radius of a cluster 
\citep{2015ApJ...806..101H, 2015ApJ...810...36M, 2017ApJ...843..128R, 2020arXiv201005920B}.  
At this radius, the density profiles of clusters change
\citep{2015ApJ...810...36M, 2020arXiv201005920B}.
However, as we do not have direct information about galaxy orbits 
in clusters  from observations (as can be obtained from simulations), we call this radius $R_{\mathrm{cl}}$.
How the virial radius $R_{\mathrm{vir}}$ and $R_{\mathrm{cl}}$ are related depends on the
structure and merging history of a cluster \citep{2015ApJ...810...36M}.
To mark the early infall zone in the PPS diagram, 
we applied Eq.~\ref{eq:infalltime}, and used $R_{\mathrm{vir}}$ in calculations.
We additionally show in the PPS diagrams  
the region, which is more likely to also include galaxies with late infall
times. The borderline of this region is  calculated with Eq.~\ref{eq:infalltime}
and using $R_{\mathrm{cl}}$.
We also used additional information
about galaxies (their star formation properties) to separate possible
early and late infallers, members of infalling components, and possible interlopes.

\subsection{Connectivity of clusters} 
\label{sect:con} 

The connectivity of a cluster  is defined 
as the number of filaments connected to a cluster
\citep[see e.g.  ][]{2000PhRvL..85.5515C, 2018MNRAS.479..973C}. 
The connectivity characterises  the growth of 
clusters and  depends on their mass and richness.
The connectivity of poor clusters is lower than that 
for rich clusters (and groups) \citep{2019MNRAS.489.5695D,
2020MNRAS.491.4294K, 2020A&A...635A.195G}.
The determination of filaments connected to a cluster may not be
easy. For example, in the high-density environment
of superclusters, 
filaments may not be well defined. Especially short filaments may be spurious
and are often excluded from an analysis 
\citep{2020A&A...642A..19M, 2020A&A...637A..31S, 2020A&A...639A..71K}.
In supercluster cores, clusters may be surrounded by various infalling 
structures, as discussed, for example, in the case of the cluster A2142 
in \citet{2018A&A...610A..82E, 2020A&A...641A.172E}. 
These may include infalling substructures,
groups, and filaments. 
To determine the connectivity of clusters in supercluster cores,
it is therefore straightforward to analyse all possible infalling substructures of clusters,
groups, and filaments near these clusters together.
When we determined the connectivity of clusters, we accordingly took all possible infalling structures into account.
Among all these structures, 
we separately searched for long filaments.
We analysed the distance distribution of 
these structures from the cluster centre, the distribution of their member
galaxies in the plane of the sky, in the sky - velocity plane,
and in the PPS diagram.

\subsection{Acceleration field in the supercluster} 
\label{sect:acc} 

To complement the predictions for the dynamical state and future evolution 
of the supercluster based on the spherical collapse model,  we 
calculated the acceleration field of the supercluster. 
In these calculations we treated the groups and
clusters in the supercluster as a source of acceleration and 
analysed how the acceleration field
would affect its further evolution. We wish to understand whether 
the acceleration is strong enough to overcome the Hubble expansion.
Our observational data are limited to three coordinates for each object,
the sky coordinates and the
redshift, which is a mixture of distance  and line-of-sight velocity. 
The dynamics of objects is fully calculable
with six coordinates \citep[an example of the inference of accelerations with all spatial
coordinates is shown in][]{2020MNRAS.494.3358K}. Because we do not have all six
coordinates, we must compensate for missing information by making assumptions. The classical and
most reliable assumption for this compensation is applied in the 
 spherical collapse model. In this model,
 the spherically averaged radial velocity around a system 
in the shell of radius $R$ can be written as $u = HR - v_{\mathrm{pec}}$, 
where $v_{\mathrm{H}} = HR$ is the Hubble expansion velocity and 
$v_{\mathrm{pec}}$ is 
the averaged radial peculiar velocity towards the centre of the system. 
The peculiar velocity at the turnaround $v_{\mathrm{pec}} = HR$ and $u = 0$. 
If $v_{\mathrm{pec}} < HR$, the system expands, 
and if $v_{\mathrm{pec}} > HR$, the system begins to collapse
\citep{2015A&A...581A.135G}. 
When using data of supercluster cores, deviations from this
model are expected as they are typically not spherical. 
In order to estimate the robustness of assuming spherical
collapse model, we used the acceleration field as a test. 
This test was performed to make a competing estimate by
swapping assumptions. Instead of spherical symmetry and assumable velocities, we
discarded the assumption about the spherical symmetry in an observational distribution
of groups and clusters in the supercluster. 
As the price
of not knowing  velocities, the present mean movements of groups and
clusters were taken to be zero. We assumed that
the dominant acceleration is caused by the most massive clusters in the supercluster. In
order to show their effect most strongly, we selected a coordinate system that
is located in the plane determined by the three richest clusters in the main part of 
the CB: A2065, A2089, and A0261. The
$x_2$ coordinate was determined by connecting A2089 and A2065, the $y_2$ coordinate
was defined to be perpendicular with $x_2$  and to be in the same plane as determined by these three
clusters. The third coordinate was determined to be perpendicular with both
$x_2$ and $y_2$. The zero point of these is the mid-point between these clusters.
This coordinate system was designed to keep the majority of the acceleration field in
the plane of the figures, hence to show how the system may evolve in the clearest way. 
We illustrate the coordinate system used in these calculations in Fig.~\ref{fig:coord}.
Errors in the calculation of the acceleration field come mostly from the assumption about
cluster velocities.

We calculated the acceleration field based on the distribution of the clusters in
the supercluster region. Each group or cluster had its 
Navarro-Frenk-White (NFW) profile determined from the \citet{2014A&A...566A...1T}
group catalogue, which is sufficient to estimate the acceleration field caused by
them. The acceleration field is estimated from
\begin{equation}
{\bf a}({\bf x}) = \sum\limits_i {\bf a}_{{\rm NFW},i} ({\bf x}-{\bf x_i} )
,\end{equation}
where the index $i$ sums over the groups or clusters in the SC, and ${\bf a}_{{\rm
NFW}, i(\cdot)}$ is a function that calculates the acceleration field by group $i$ at
${\bf x}$ assuming an NFW density profile. Single galaxies are taken into account as
the possible brightest members of faint groups, and as above, we used the median
mass of the poorest groups as their mass. 
Our acceleration field does not include the matter that is not bound to the clusters
as a source of the acceleration field.

\section{Results }
\label{sect:results}

In this section we present the analysis of the richest clusters in high-density cores
of the CB. For each cluster we show the sky distribution of galaxies from different populations,
the location of
galaxies from various populations in the PPS diagrams 
and the $D_n(4000)$ index versus clustercentric distance  $D_c$ plots,
and give the radii $R_{\mathrm{cl}}$ in Table~\ref{tab:clfil}.  
We also show the sky distribution of galaxies in filaments that begin near clusters.
We briefly analyse the location of galaxies from various populations,  and compare
galaxy populations in clusters and in their close environment.
We summarise the results for the clusters in Sect.~\ref{sect:clsummary}.

\subsection{Cluster A2065}
\label{sect:a2065}  

\begin{figure*}[ht]
\centering
\resizebox{0.44\textwidth}{!}{\includegraphics[angle=0]{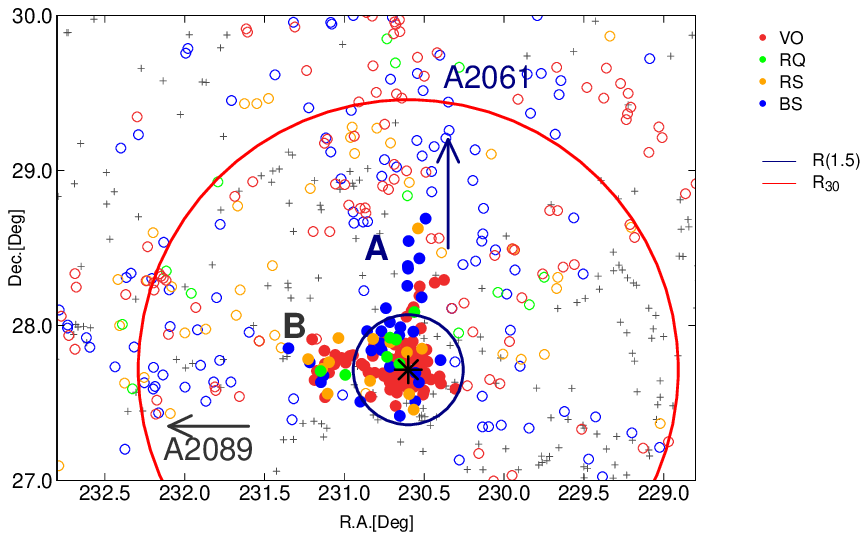}}
\resizebox{0.38\textwidth}{!}{\includegraphics[angle=0]{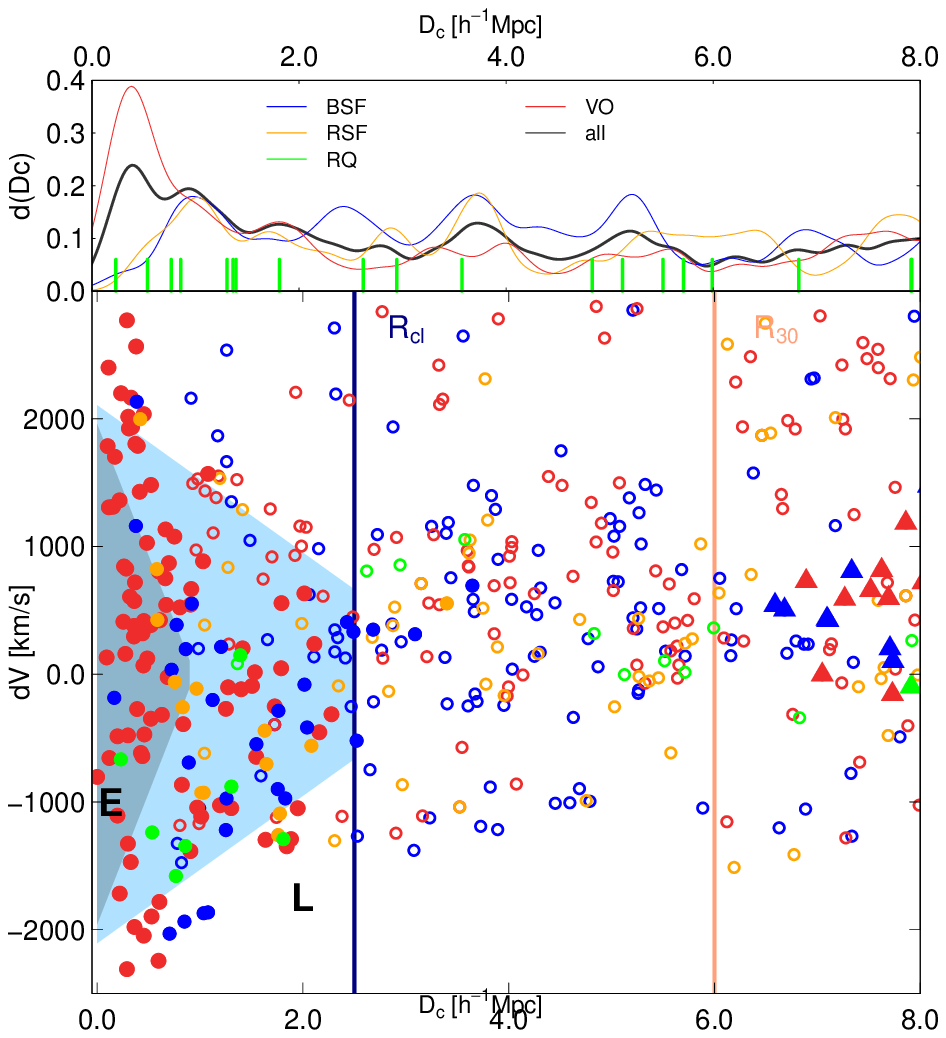}}
\resizebox{0.44\textwidth}{!}{\includegraphics[angle=0]{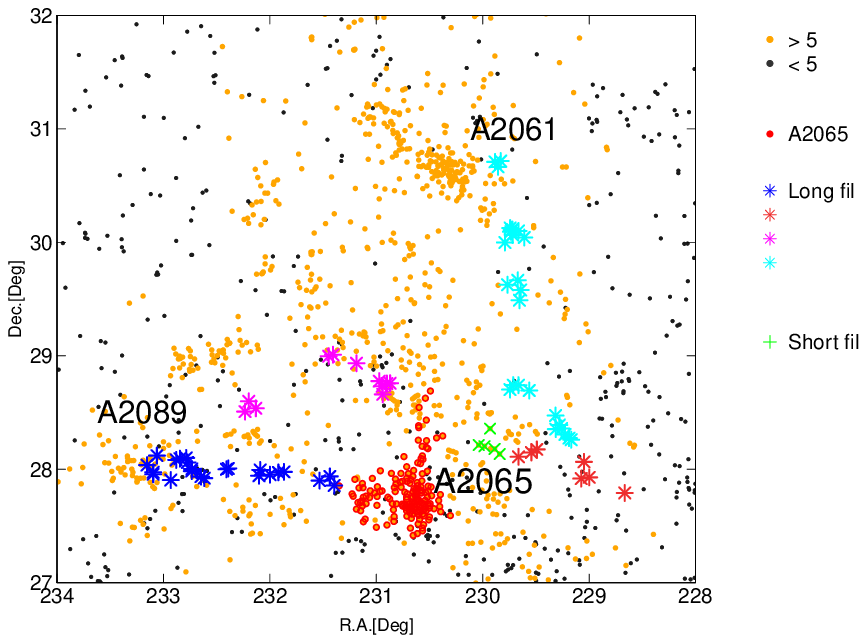}}
\resizebox{0.37\textwidth}{!}{\includegraphics[angle=0]{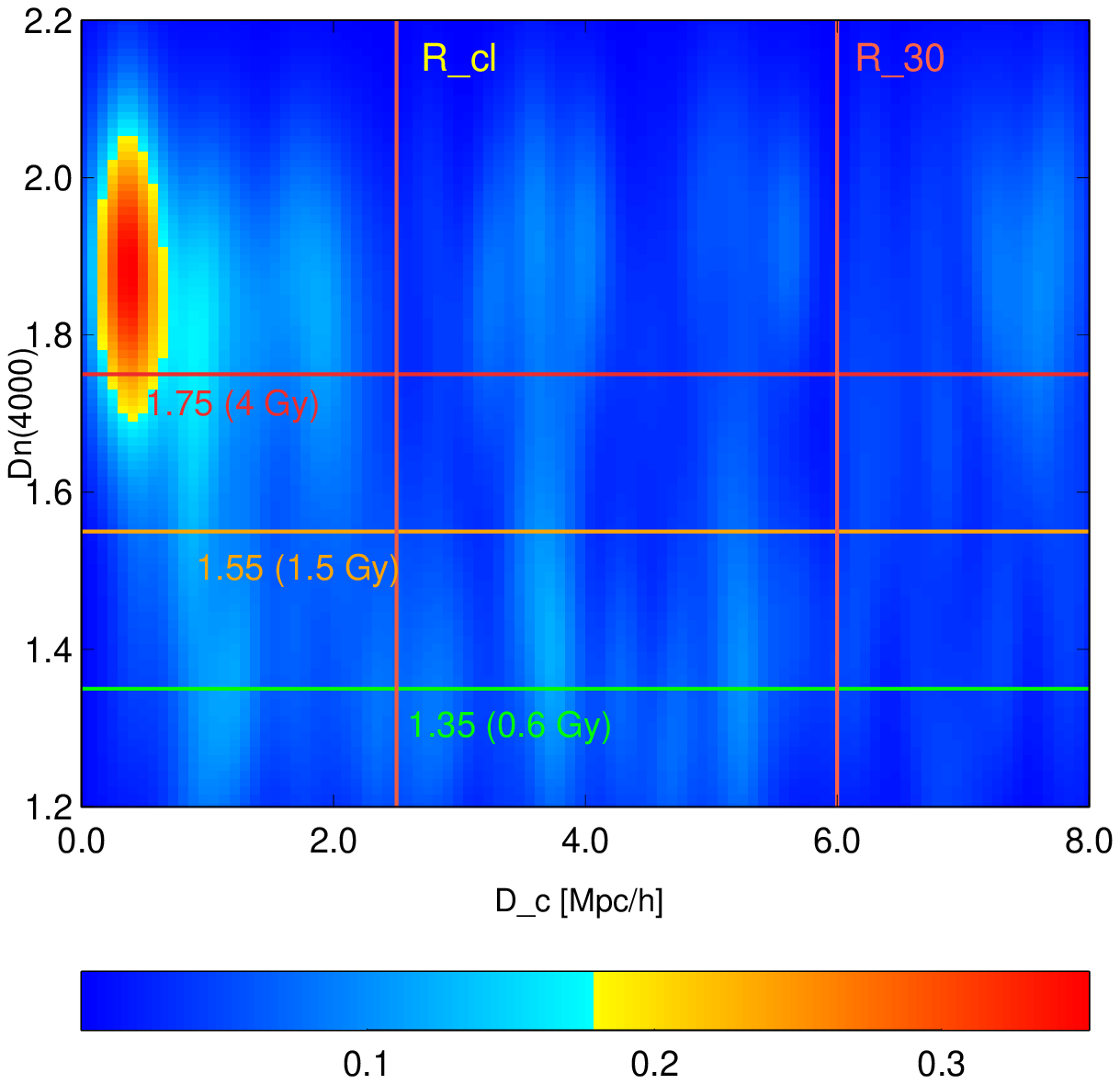}}
\caption{
Left panels:Distribution of galaxies in A2065 in the plane of the sky. The colours in the upper panel correspond
to galaxies from different populations (BSF - blue, RSF - orange, RQ - green,
and VO - dark red). Filled symbols denote galaxies in different components of the cluster. 
Empty circles show galaxies that are not cluster members. Grey crosses
denote galaxies in the low-density region with $D8 < 5$.  
The inner black circle indicates the scale with a radius $1.5$~\Mpc. 
The outer red circle corresponds to the radius $R_{30}$ ($= 6$~\Mpc\ for A2065). 
A and B denote substructures that extend from the cluster (see text).
The colour in the lower panel shows galaxies in long filaments with a length larger than 5~\Mpc,
and in green we plot galaxies in a short filament with a length of 3~\Mpc, which
connects the long filament with the cluster.
The upper right panel shows the PPS diagram and the distribution of the clustercentric distances $D_c$
for galaxies from different populations in A2065 and in its environment towards A2089.
Filled circles represent A2065 members, and filled triangles (at clustercentric
distances $D_c > 6$~\Mpc) show A2089 members. 
Empty symbols mark galaxies that are not members of these two clusters.
The blue region shows the early infall (E) region, and light blue shows late infall (L) region. 
The lower right panel presents the $D_n(4000)$ index versus clustercentric distance  $D_c$
for galaxies from A2065 to A2089. 
Colours show the density of points at a given location in the plot.
Vertical lines correspond to the infall zone of the cluster, $R_{\mathrm{cl}}$, and
to the radius $R_{30}$.  
Horizontal lines show $D_n(4000)$ index values
$D_n(4000) = 1.75$, $D_n(4000) = 1.55$, and $D_n(4000) = 1.35$ 
(see Sect.~\ref{sect:galpop}). 
}
\label{fig:a2065radecpps}
\end{figure*}

We start our analysis from the richest cluster in the CB, cluster A2065.
In Fig.~\ref{fig:a2065radecpps} 
(upper left panel) we plot the sky distribution of galaxies in the region of A2065.
Here galaxies from different populations
are marked with different symbols. To show the physical scale of the cluster,
we add to Fig.~\ref{fig:a2065radecpps} 
(upper left panel)  a circle with radius of $1.5$~\Mpc. We also  
plot a circle with radius $R_{\mathrm{30}}$; we explain below 
why we use this notation. 
We show the distribution of
galaxies from various populations in A2065 and around it using the PPS diagram
and the $D_n(4000)$ index versus clustercentric distance  $D_c$ plot 
(Fig.~\ref{fig:a2065radecpps}, right panels).
We analyse possible infalling structures and filaments near the cluster
to determine its connectivity (Fig.~\ref{fig:a2065radecpps}, lower left panel).

The sky distribution of galaxies at A2065 (Fig.~\ref{fig:a2065radecpps},
upper left panel)
shows that this cluster is located at the intersection of galaxy systems connecting it 
with clusters A2061 in the north and with A2089 in the east.
Substructures appear to extend from the cluster in both directions.
The substructure analysis of A2065 revealed four components in the cluster.
The first component is the cluster itself.
Another component lies along the line of sight;
in the sky projection, it is projected onto  the main component. 
The other two components of A2065 are elongated and point towards clusters A2061 and A2089
(Fig.~\ref{fig:a2065radecpps}, upper left panel, denoted A and B).

Next we analysed the galaxy populations in and around A2065 using PPS diagrams.
To avoid contamination due to projections, we calculated the  PPS diagrams
separately for the direction to clusters A2089 and A2061. To find the PPS diagram 
from A2065 to A2089, we used galaxies in the declination interval $26.8 - 28.8$~degrees
(Fig.~\ref{fig:a2065radecpps}, upper right panel).
In order to not to overcrowd the text with figures, we do not show 
the PPS diagram for galaxies in the direction of A2061. 
In the upper panel of the PPS diagram, we show the distribution of clustercentric distances
of galaxies from different populations.
We complement the PPS diagram with a figure in which we plot the $D_n(4000)$ index
of galaxies versus clustercentric distances of galaxies, $D_c$,
in the same region as shown in the PPS diagram (Fig.~\ref{fig:a2065radecpps}, lower right panel).

In the  PPS diagram  in the infall region of the line-of-sight component 
there are some blue star-forming and recently quenched galaxies, and also some VO
galaxies  (Fig.~\ref{fig:a2065radecpps}, upper right panel).
The presence of star-forming and recently quenched galaxies is a signature
that this component  is probably a remnant of a group that is falling into the cluster. 
This conclusion is supported by X-ray and radio studies \citep{1999ApJ...521..526M, 2013ApJ...779..189F}.
\citet{1999ApJ...521..526M} proposed using an analysis of X-ray data that 
 a late-stage merger occurs here, perhaps well after a core
passage. From the analysis of XMM-Newton observations data,
\citet{2005AdSpR..36..630B} concluded that the data show signatures 
of an ongoing merger. 
\citet{2013ApJ...779..189F} detected diffuse radio emission,
which they identified as possible giant radio halo at the
location of X-ray peak.
We can also say this in another way: X-ray and radio studies
have revealed a possible infalling group in A2065 that coincides with the line-of-sight component detected in A2065.
Some galaxies of this component are recently quenched,
they are visible on the PPS diagram as galaxies with high velocities and small
clustercentric distances. Their quenching may be related to their infall into the cluster.

Comparison of galaxy populations
in the main cluster and in the components shows that the main cluster is
populated mostly by VO galaxies, 
as also shown in Fig.~\ref{fig:a2065radecpps} (lower right panel, 
$D_n(4000)$ index versus clustercentric distance  $D_c$ plot). 
Star-forming galaxies, both red and blue, populate 
the region 
between the early infall region and the boundaries of the 
the cluster, up to $R_{\mathrm{cl}}$~\Mpc. They may belong to late infallers,
or they may be falling into the cluster now and not yet be quenched in the 
cluster environment. Some of these galaxies may be interlopers.
The upper panel of the PPS diagram shows an excess of blue star-forming
galaxies at $R_{\mathrm{cl}}$, followed by  a lower excess of red star-forming
galaxies. Their star formation may be triggered by the infall into the cluster. 
At very small clustercentric distances lie
VO galaxies with high peculiar velocities.
These galaxies might belong to the
part of an infalling cluster after a core passage, as suggested by
X-ray data \citep{1999ApJ...521..526M}. RQ and star-forming galaxies may be at their first infall.

In the PPS diagram, galaxies with $D_c > R_{\mathrm{cl}}$
(up to clustercentric distances
$D_c \approx 6$~\Mpc)
are not yet members of the (main) cluster. They either belong to substructures
or are falling into the cluster.
Galaxies at this clustercentric distance interval with very large 
velocity differences may be there due to projection effects and 
do not belong to the infalling population. 
At clustercentric distances up to 
$D_c \approx 6$~\Mpc,\ individual maxima in the 
clustercentric distance distribution correspond to small groups and filaments
near the cluster (we discuss them below). At clustercentric distances in the range
$D_c \approx 6-7$~\Mpc,\ there is a minimum in the distance distribution of
galaxies. This distance interval is populated by only a few 
star-forming galaxies. At higher clustercentric distances,
close to cluster A2089, the number of galaxies increases again.
In the PPS diagram, the radius $D_c \approx 6$~\Mpc\ marks the borders of the late infall region.
We may assume that this radius 
marks the borders of the possible sphere of influence of cluster
A2065.  We denote the clustercentric distance at which the minimum occurs
as the radius $R_{30}$ and give the reasons of this notation
in Sect.~\ref{sect:discussion}.

To determine the connectivity of A2065, we 
identified galaxy filaments within the sphere of influence of A2065,
and galaxy groups and galaxies belonging to filaments. Then we 
analysed their properties and their
location in the plane of the sky and in the 
PPS diagram. Filaments near clusters may be poorly defined by an automatic algorithm, and sometimes
elongated groups may be included in a filament catalogue. Therefore we checked all the 
systems near the cluster, and excluded very short, spurious 
filaments with lengths shorter than $3$~\Mpc. The mean velocity dispersion
of groups is $\approx 200$~km/s, and our automatic
algorithm may confuse these groups with short filaments. 
Typically, filaments with a length 
in the range of $3-5$~\Mpc\ were associated with a group and a few galaxies near it.
Filaments with lengths greater than $5$~\Mpc\  either connect two clusters
(A2065 and A2089, or A2065 and A2061)
or  enter the low-density region around the supercluster.

As mentioned, A2065 has three infalling substructures. As in the case of cluster A2142
in \citet{2020A&A...641A.172E}, they were taken into account to determine the
connectivity of A2065. In the region of influence of the cluster, two groups are associated with short filaments. 
Four filaments with length $> 8 $~\Mpc\ are also connected to A2065.
One of these filaments connects clusters A2065 and A2089, another
extends from A2065 to A2061.
Two long filaments continue in the low-density region.
When we count them all, the total connectivity $\pazocal{C}$ of A2065
is $\pazocal{C} = 9$. In Table~\ref{tab:clfil} we list the total number
of different systems near the cluster and the number of long filaments.

In the lower right panel of Fig.~\ref{fig:a2065radecpps} we show the
value of the $D_n(4000)$ index versus clustercentric distance  $D_c$. This
figure shows that VO galaxies populate the
early infall region of A2065. This figure clearly shows that
galaxies that probably have been falling
into the cluster during its formation stopped their star formation and became VO
galaxies in the cluster
environment. Blue star-forming galaxies lie in the infall zone of A2065. 
The infall zone is also populated by galaxies 
in transition (RQ and RSF galaxies).
The gap in the galaxy distribution at $R_{30}$ is also visible, although this is not strong. 
This means that galaxy groups and filaments that extend from  A2065 to
A2089 are not continuous. In other words, they do not touch, but rather greet 
each other across a distance\footnote{
We could call this "elbow bump" or 
"the Corona handshake". 
It will remind us this period of time when, due to the Covid-19 restrictions,
we worked separately at homes without close contacts between us while preparing this study 
}.

In the A2065 sphere of influence,
there is a local excess of red star forming
galaxies at $D_c \approx 3$~\Mpc, and another local excess of
blue star-forming galaxies at $D_c \approx 4$~\Mpc. We might witness the transformations of star-forming galaxies,
in which at first the colour of galaxies changes.

\subsection{A2061 to A2067 region}
\label{sect:a2061}  

\begin{figure*}
\centering
\resizebox{0.44\textwidth}{!}{\includegraphics[angle=0]{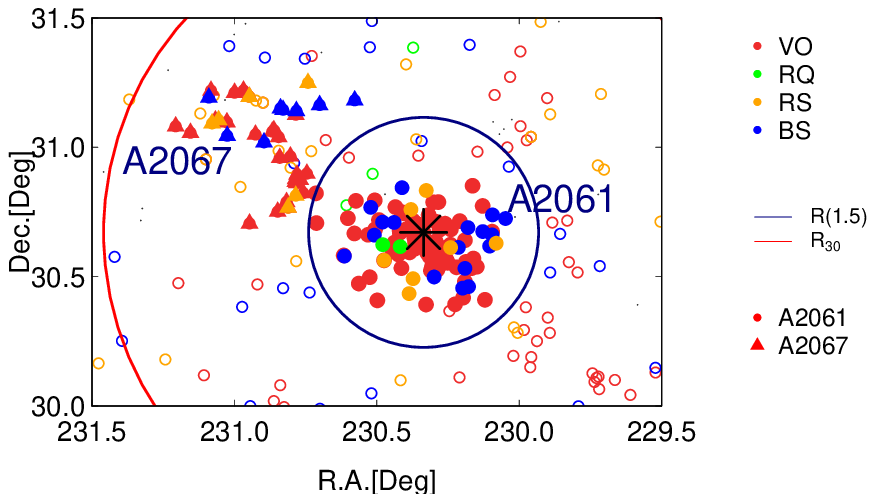}}
\resizebox{0.38\textwidth}{!}{\includegraphics[angle=0]{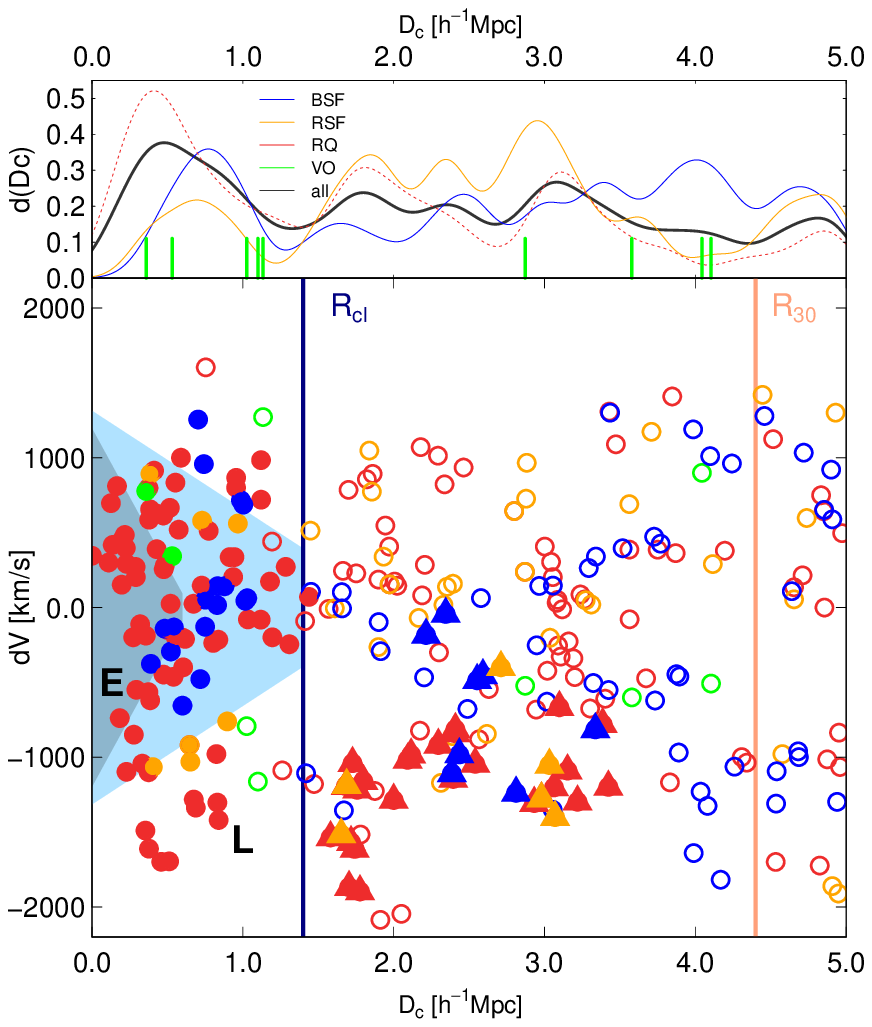}}
\resizebox{0.44\textwidth}{!}{\includegraphics[angle=0]{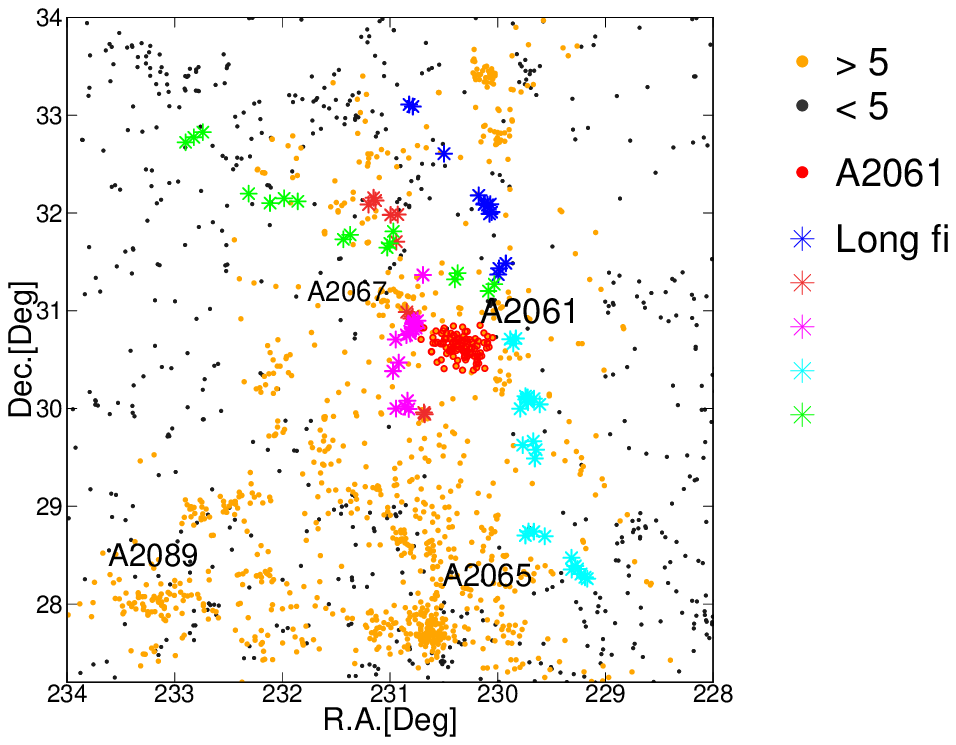}}
\resizebox{0.37\textwidth}{!}{\includegraphics[angle=0]{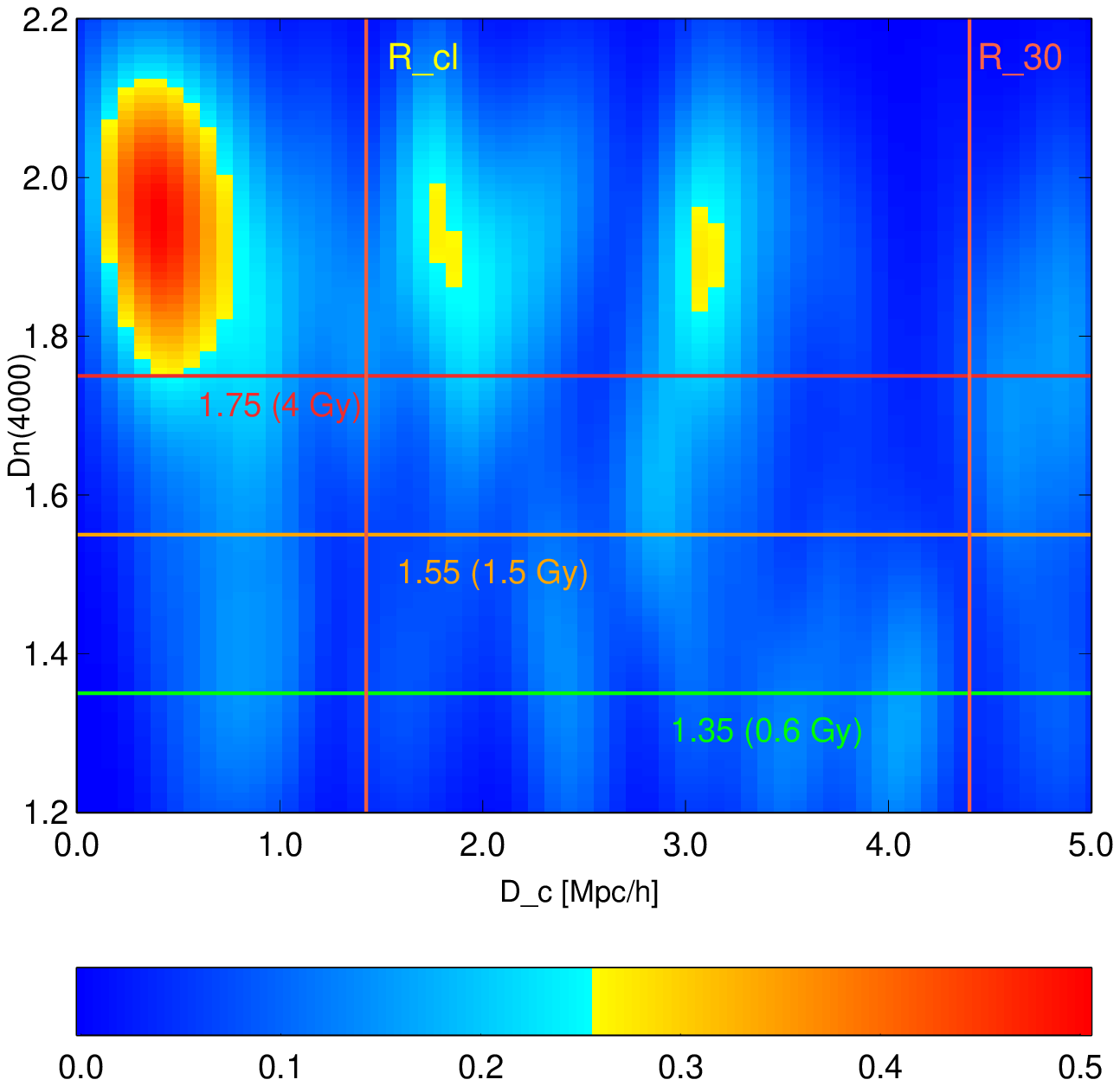}}
\caption{
Same as in Fig.~\ref{fig:a2061radecpps} for A2061. Triangles in the upper panels
denote galaxies from cluster A2067.
}
\label{fig:a2061radecpps}
\end{figure*}

We show the distribution in the sky of galaxies in cluster A2061 in 
the left panel of Fig.~\ref{fig:a2061radecpps}.
The PPS diagram up to clustercentric distances $5$~\Mpc\ and the 
$D_n(4000)$ index versus $D_c$
are given in the right panels of Fig.~\ref{fig:a2061radecpps}. 
Here galaxies from the low-mass cluster 
A2067 near A2061 are plotted with separate symbols. These figures show that clusters A2061 and 
A2067 form a close pair;  in the distribution in the sky,
they almost overlap. In the PPS diagram, galaxies from A2067
lie in the infall region of A2061. Interestingly, the galaxies from A2067 that are 
closest to A2061 all contain very old stellar populations, and they also
include two red star-forming galaxies. 
There are also other RSF galaxies in both A2061 and A2067.
In A2061, they may be just falling in, as they lie in and near boundaries of
the late infall (light blue) zone (as seen also in the lower right panel of
Fig.~\ref{fig:a2061radecpps}). 
The virialised part of the PPS diagram is populated mostly by VO galaxies.
Blue star-forming galaxies lie in the 
late infall zone 
of A2067, giving the impression that this cluster may consist
of two merging groups. These galaxies cause a maximum in the clustercentric
distance distribution of BSF galaxies between $R_{\mathrm{vir}}$ and $R_{\mathrm{cl}}$
(upper panel of the PPS diagram).
We might therefore witness several
simultaneous mergers: two merging groups that form A2067, which in turn is falling
into A2061. 
The upper panel of the PPS diagram shows a minimum in the galaxy distribution around A2061,
followed by a small maximum, especially  in the distribution of RFS galaxies.
Some of these galaxies are members of A2067, and some of them belong to other groups.
They include VO galaxies
and also star-forming galaxies (Fig.~\ref{fig:a2061radecpps}, lower right panel).
{\it mclust} identified three components in A2061, which may partly be due to its
elongated shape.

Both X-ray and radio observations have demonstrated complicated and structure-rich
X-ray and radio emission in the area of A2061 and A2067
\citep[][and references therein]{2013ApJ...779..189F, 2004MNRAS.353.1219M}. 
\citet{2013ApJ...779..189F} showed that 
A2061 is elongated and has an elongated
radio relict that form a diagonal from north-east to 
south-west in Fig.~\ref{fig:a2061radecpps}. \citet{2013ApJ...779..189F} discussed the 
possibility that a filament of galaxies might connect A2061 and A2067. 
This filament is formed by the same galaxies that connect these clusters in
Fig.~\ref{fig:a2061radecpps}. \citet{2004MNRAS.353.1219M} found from X-ray data
that a group of galaxies may have been falling into A2061 along the axis that
connects A2061 and A2067. Thus it is possible that several merging events in the
past and at present have been occurred along this axis, and in the future, the whole cluster
A2067 will merge with A2061. 
These mergers may affect the star-forming properties of galaxies in both clusters.
We plan to study  the A2061 and A2067 cluster pair in more detail in the future.

There is a minimum in the distance distribution of galaxies around A2061 at approximately $4$~\Mpc\ (Fig.~\ref{fig:a2061radecpps}, right panels).
This radius, $R_{30}$, defines the radius of the sphere of influence for A2061.
Galaxies from A2067 are within this sphere.
Near $R_{30}$, at clustercentric distances of approximately $3.2 < D_c < 4$~\Mpc,\
most galaxies are with very young stellar populations 
(Fig.~\ref{fig:a2061radecpps}, right panels). 

The analysis of groups and filaments in the sphere of influence of A2061
showed that this cluster has three short filaments near it, and one of them connects
A2061 and A2065. All these short filaments are associated with infalling groups
in the PPS diagram. A2061 also has five long filaments.
One of them connects A2061 and A2065 and is related with the radio
relict mentioned above \citep{2013ApJ...779..189F}.
Another filament is directed towards
A2065 but does not extend to it. Three filaments that are
longer than $10$~\Mpc\ are directed towards the low-density region
around the supercluster. We thus obtain for A2061 a total 
connectivity $\pazocal{C} = 8$, with five long filaments (Table~\ref{tab:cl}).

\begin{figure*}
\centering
\resizebox{0.44\textwidth}{!}{\includegraphics[angle=0]{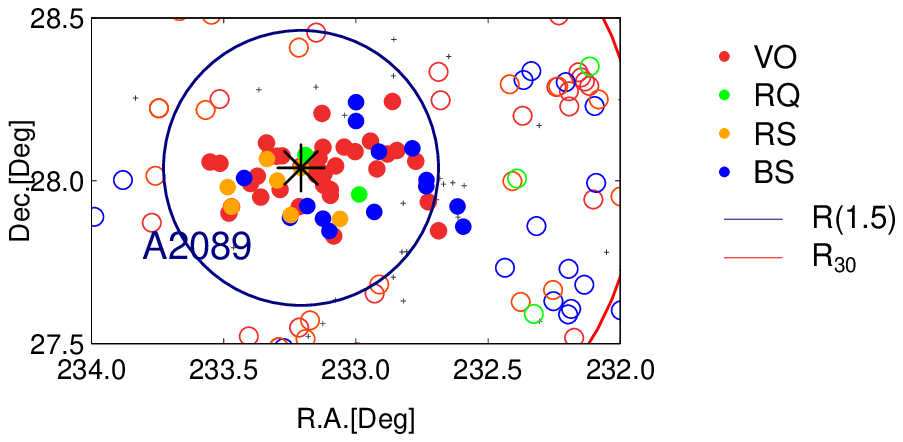}}
\resizebox{0.38\textwidth}{!}{\includegraphics[angle=0]{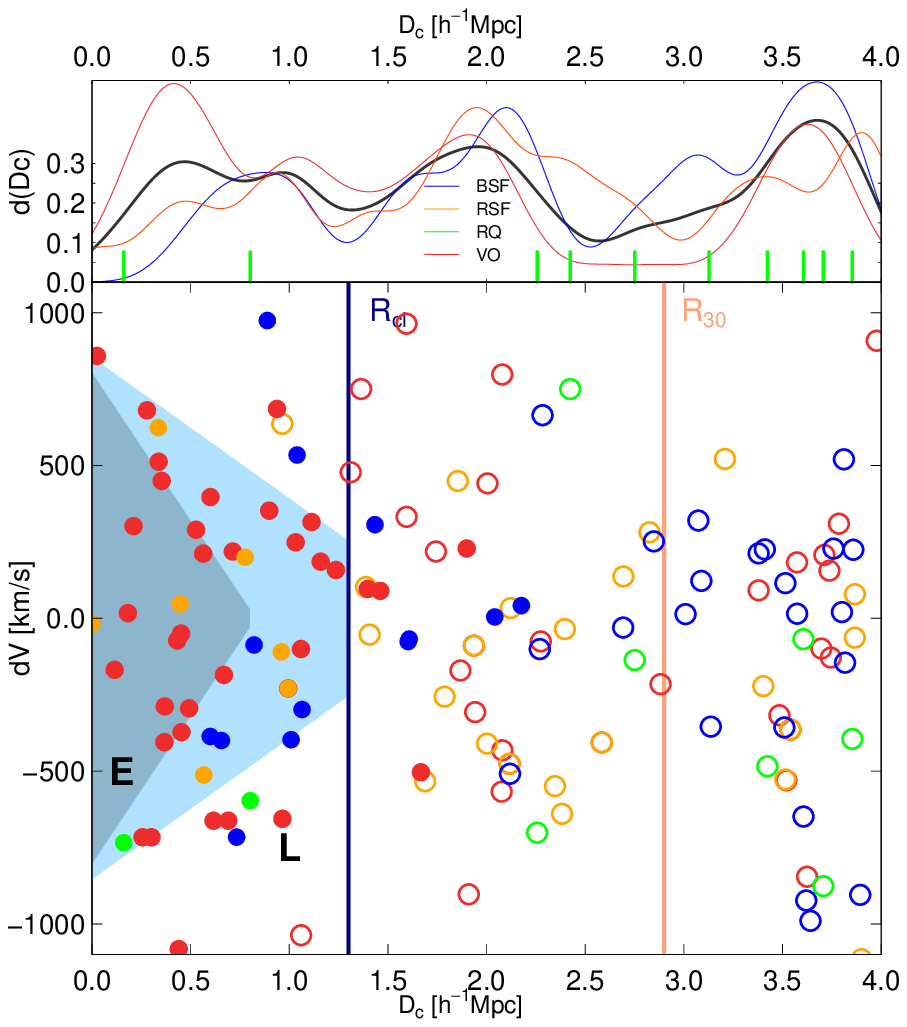}}
\resizebox{0.44\textwidth}{!}{\includegraphics[angle=0]{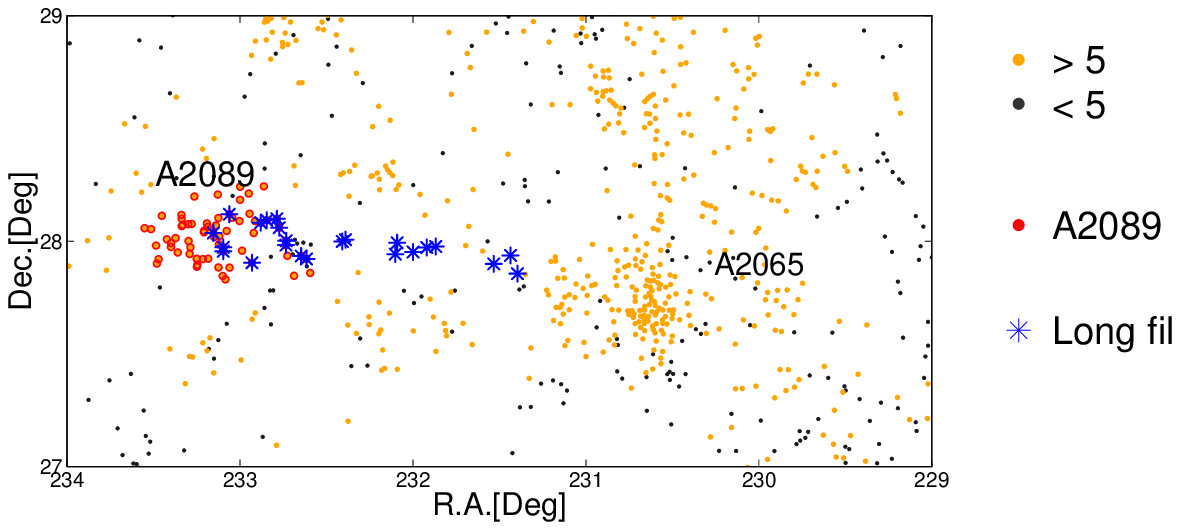}}
\resizebox{0.37\textwidth}{!}{\includegraphics[angle=0]{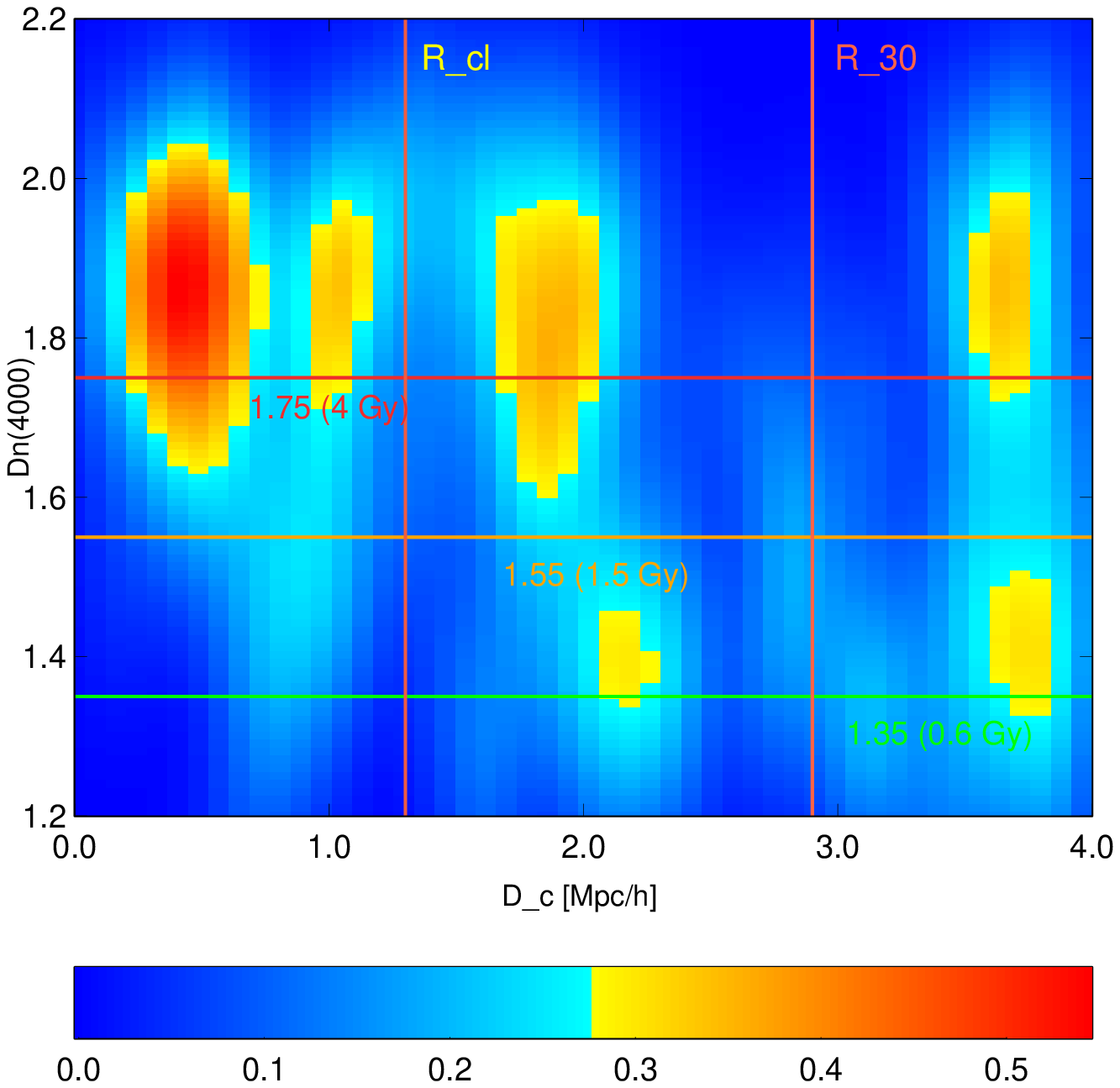}}
\caption{
Same as in Fig.~\ref{fig:a2061radecpps} for A2089.
}
\label{fig:a2089radecpps}
\end{figure*}

\subsection{A2089 region}
\label{sect:a2089}

Cluster A2089 in Fig.~\ref{fig:a2089radecpps} 
is elongated along the direction towards A2065. It has a tail of galaxies pointing towards A2065. 
However, normal mixture modelling showed that A2089
is a one-component cluster. \citet{2007MNRAS.379.1011A} found that 
the galactic planes of the galaxies in A2089 tend
to lie in the plane defined by the elongated shape of the cluster
(and filaments, as our study shows).
The central part of the cluster (early infall zone in the PPS diagram)
is populated mostly by VO galaxies; there are blue and red star-forming galaxies 
in the late infall  (light blue)  zone and in the tail. Two RQ
galaxies lie in the late infall zone of A2089.
We may assume that the star formation properties of these galaxies 
are affected by the infall into the cluster.
There are also galaxies with clustercentric distances smaller than $R_{\mathrm{cl}}$
and large velocity differences between these galaxies and the cluster centre.
Some of them may be interlopers (due to their high velocities), but an
RQ galaxy, for instance, may be falling into the cluster, and its star formation may be recently
quenched because of this. 
Figure~\ref{fig:a2089radecpps} (upper right panel) shows
a minimum in galaxy distribution at $R_{\mathrm{cl}}$, and another
broad minimum in the clustercentric distance distribution at approximately
$2.5 - 3$~\Mpc, where only a few BSF galaxies lie. 
In Fig.~\ref{fig:a2089radecpps}, the upper panel of the PPS diagram and the lower
panel ($D_n(4000)$ index versus clustercentric distance  $D_c$ plot) show that
of the infalling galaxies at $D_c > R_{\mathrm{cl}}$, the VO galaxies
are closer to the cluster than the star-forming galaxies. It is possible that the star formation
of these galaxies has been quenched because of the infall, but this may also be
evidence of preprocessing of galaxies in groups before infall to the cluster.
The overall density minimum at this clustercentric distance interval suggests that
this distance defines the size of the sphere of influence for A2089, $R_{30}$.

A2089 has only one infalling group and one outgoing filament. This
filament extends from A2089 to A2065. According to the filament finder,  
galaxies in the tail that extends from A2089
also belong to this filament. 
Therefore the total connectivity of A2089 is only $\pazocal{C} = 2$, with one long
filament (Table~\ref{tab:cl}).

\subsection{Gr2064 region}
\label{sect:gr2064}  

\begin{figure*}[ht]
\centering
\resizebox{0.44\textwidth}{!}{\includegraphics[angle=0]{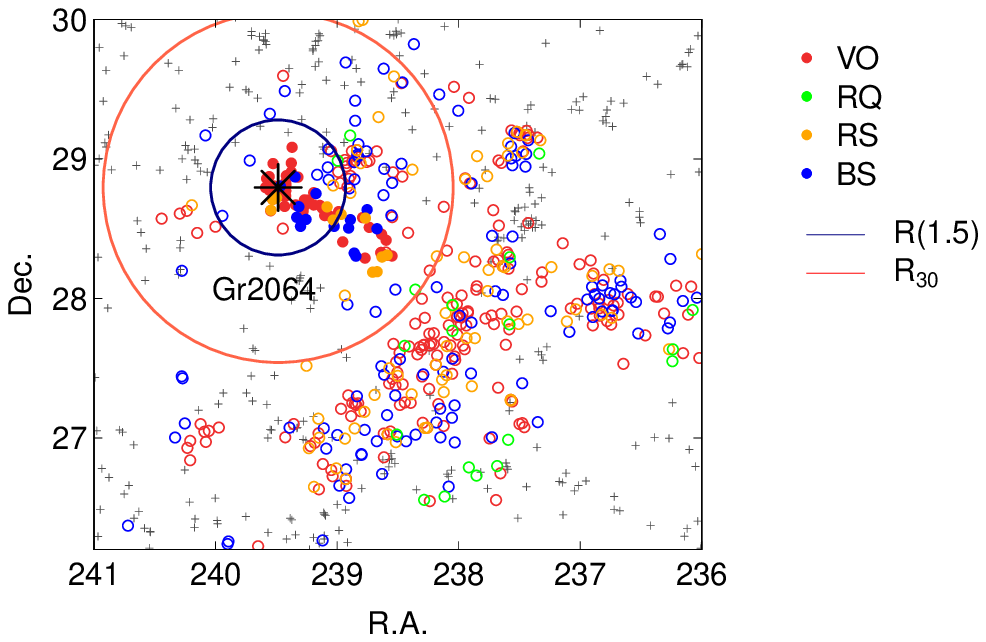}}
\resizebox{0.38\textwidth}{!}{\includegraphics[angle=0]{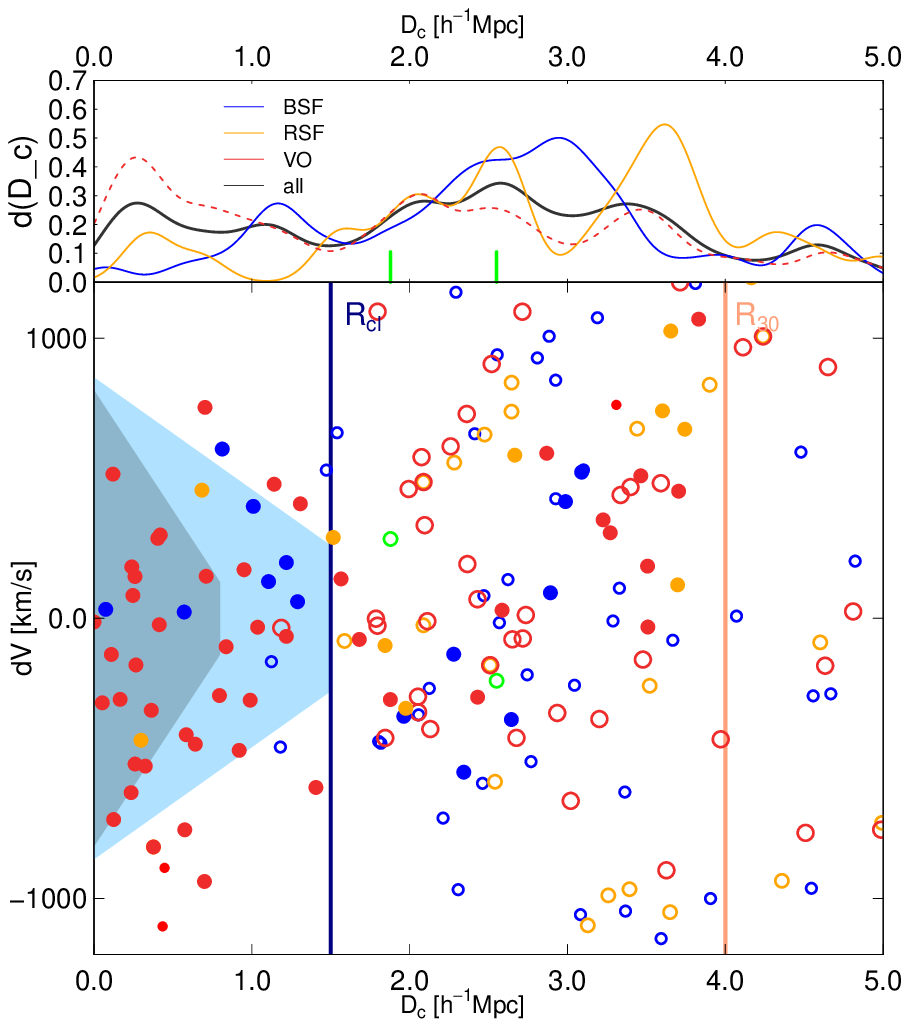}}
\resizebox{0.44\textwidth}{!}{\includegraphics[angle=0]{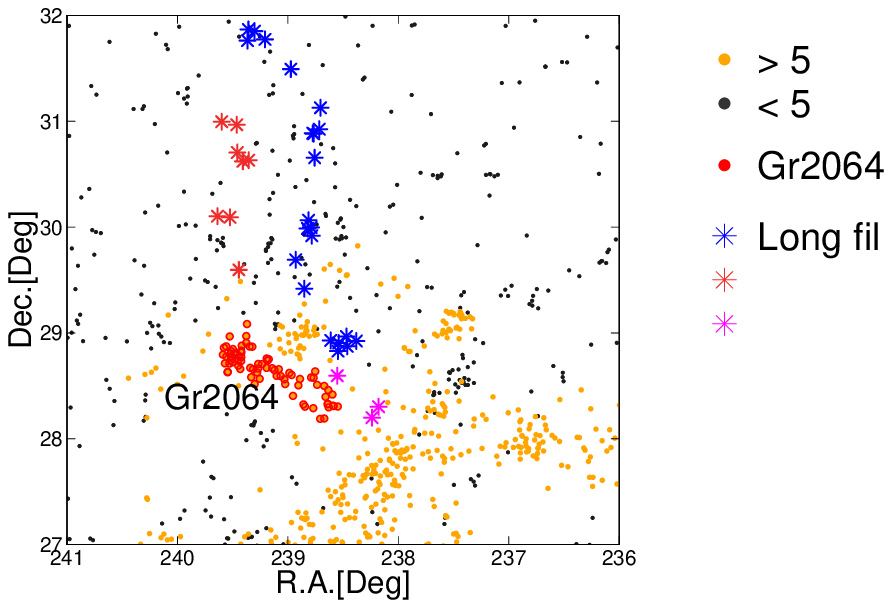}}
\resizebox{0.37\textwidth}{!}{\includegraphics[angle=0]{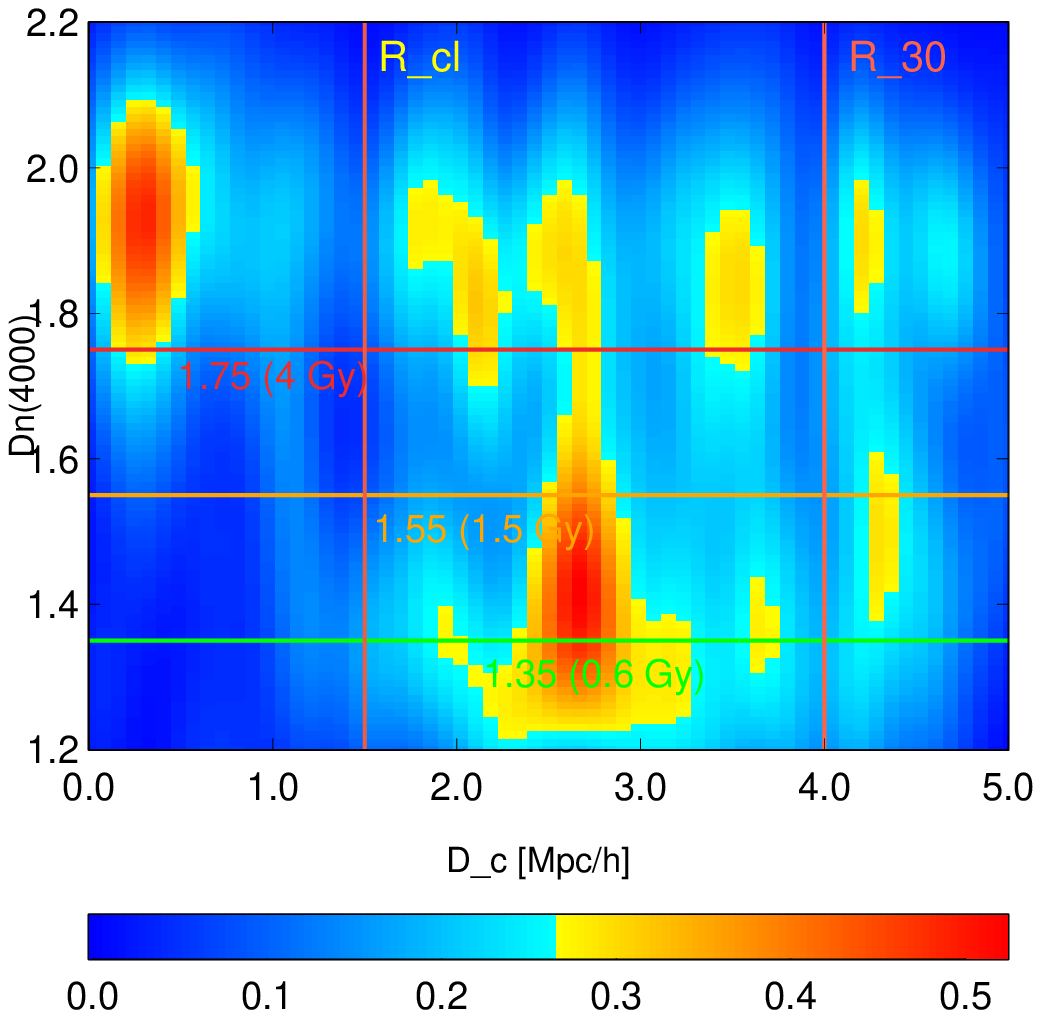}}
\caption{
Same as in Fig.~\ref{fig:a2061radecpps} for Gr2064.
}
\label{fig:gr2064radecpps}
\end{figure*}

Figure~\ref{fig:gr2064radecpps} shows galaxies in  cluster Gr2064 in the plane of the sky,
in the PPS diagram, and in the $D_n(4000)$ index versus $D_c$ plot.
Gr2064 is a multicomponent cluster, as shown in the results of normal mixture modelling.
Gr2064 consists of the main cluster 
and two elongated
substructures pointing towards neighbouring structures  (Fig.~\ref{fig:gr2064radecpps}, 
upper right panel).
These substructures extend from the cluster up to a distance of approximately $4$~\Mpc.
There is a rich group of 30 galaxies near Gr2064, and as the PPS diagram suggests,
this group is falling into Gr2064. It may have already started to merge; the nearest galaxies from
this group are closer than $1.5$~\Mpc\ to the centre of Gr2064.  
The right  panels of Fig.~\ref{fig:gr2064radecpps}  show an excess of 
star-forming galaxies  (blue, with very young galaxy populations, and RSF galaxies) 
in the late infall region of the cluster
at clustercentric distances $D_c > R_{\mathrm{vir}}$~\Mpc. 
One recently quenched galaxy also lies in this region. No rich group near other clusters 
in the CB has such a high fraction of star-forming galaxies 
(Figs.~\ref{fig:a2065radecpps} - \ref{fig:a2089radecpps}; in Sect.~\ref{sect:comparison}
we compare galaxy populations in and around clusters in more detail).
A minimum in galaxy distribution
at clustercentric distances $\approx 4$~\Mpc\ is visible in the galaxy distribution
in the sky and in the PPS diagram. Only a few (red and blue)
star-forming galaxies lie there. This distance determines the size of the sphere
of influence for Gr2064.

Cluster Gr2064 has two elongated substructures and one infalling group.
Three long filaments start in the sphere of influence of Gr2064. Two of these
extend in the direction of the low-density region around the cluster; they 
are longer than $10$~\Mpc. A filament with a length of $5$~\Mpc\  connects Gr2064 with the other nearby cluster, as shown in Fig.~\ref{fig:gr2064radecpps}.
Thus the total connectivity of Gr2064 is $\pazocal{C} = 6$, with three long filaments.

\subsection{Galaxy populations in clusters and in the spheres of influence}
\label{sect:comparison}

\begin{table*}[ht]
\caption{Galaxy populations as used in this paper.}
\begin{tabular}{lll} 
\hline\hline  
(1)&(2)&(3)\\      
\hline 
Population & Abbr. &Definition  \\
\hline                                                    
Blue star-forming galaxies  & BSF &  $(g - r)_0 < 0.7$, $\log \mathrm{SFR} \geq -0.5$       \\
Red star-forming galaxies   & RSF &  $(g - r)_0 \geq 0.7$, $\log \mathrm{SFR} \geq -0.5$    \\
Recently quenched galaxies  & RQ  &  $D_n(4000) \leq 1.55$,  $\log \mathrm{SFR} < -0.5$    \\
Galaxies with very old stellar populations & VO &  $D_n(4000) \geq 1.75$  \\
\hline
\label{tab:galpops}  
\end{tabular}\\
\tablefoot{                                                                                 
Columns are as follows:
(1): Galaxy population;
(2): Abbreviation;
(3): Definition of a given population.
}
\end{table*}

In this section we compare galaxy populations in four clusters of the CB supercluster 
and in cluster A2142. 
The supercluster SCl~A2142 has a high-density core with 
 a clear gap in the galaxy distribution, which defines the boundary of the core.
Within the high-density core, all galaxies and groups are falling into cluster A2142, thus 
it corresponds to the sphere of influence of A2142.
Its radius is $\approx 5$~\Mpc\ \citep{2020A&A...641A.172E}.
Therefore we also compared the galaxy populations in clusters
and in the spheres of influence of the clusters from the CB
and in A2142. 
In the comparison we used data
of the $D_n(4000)$ index and stellar masses $\log M^{\mathrm{*}}$ of galaxies 
introduced in Sect.~\ref{sect:galpop}. The
$D_n(4000)$ index can be used to characterise the age of galactic stellar populations
and star formation rates.
The value $D_n(4000) = 1.55$ corresponds to 
a mean age of stellar populations of about $1.5$~Gyr.
We used this limit to separate galaxies with old and young
stellar populations. Galaxies with young stellar populations have  
$D_n(4000) \leq 1.55$. 
The value $D_n(4000) \geq 1.75$ corresponds to 
a mean age of stellar populations of about $4$~Gyr or older.
The value $D_n(4000) = 1.35$ limits galaxies with very young stellar populations
with a mean age of only approximately  $0.6$~Gyr (see references in Sect.~\ref{sect:galpop}).
The galaxy populations are summarised in Table~\ref{tab:galpopdef},
and we briefly show them also in Table~\ref{tab:galpops}.

The galaxy content in cluster A2142 and in supercluster SCl~A2142 was analysed
in \citet{2018A&A...610A..82E} and in \citet{2018A&A...620A.149E}, and in the
comparison, we use data from these papers. As in the present paper, 
\citet{2018A&A...610A..82E} and \citet{2018A&A...620A.149E} used SDSS data 
in their analysis of SCl~A2142 and defined galaxy populations in the same way
as they are defined here. In the following comparison we only include galaxies
from the magnitude-limited complete sample from both superclusters,  
with a magnitude limit $M_r = -19.6$~mag.
We used a Kolmogorov-Smirnov test to estimate the statistical significance 
of the differences between the galaxy populations.
We considered that the differences between distributions are 
{\it \textup{highly}} significant 
when the $p$ - value (the estimated probability of rejecting the hypothesis
that the distributions are statistically similar) $p \leq 0.01$.
We considered that the differences between distributions are significant 
when the $p$ - value $p \leq 0.05$. 

\begin{figure}[ht]
\centering
\resizebox{0.23\textwidth}{!}{\includegraphics[angle=0]{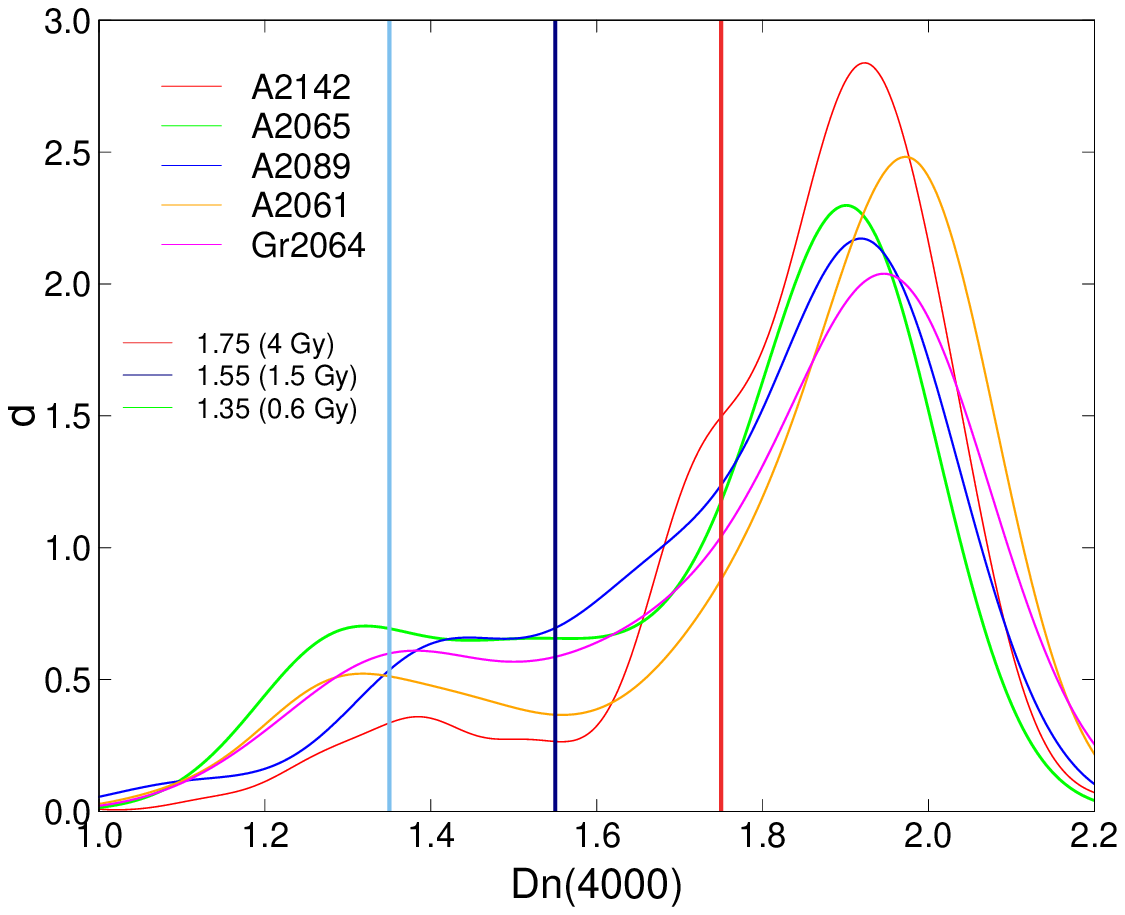}}
\resizebox{0.23\textwidth}{!}{\includegraphics[angle=0]{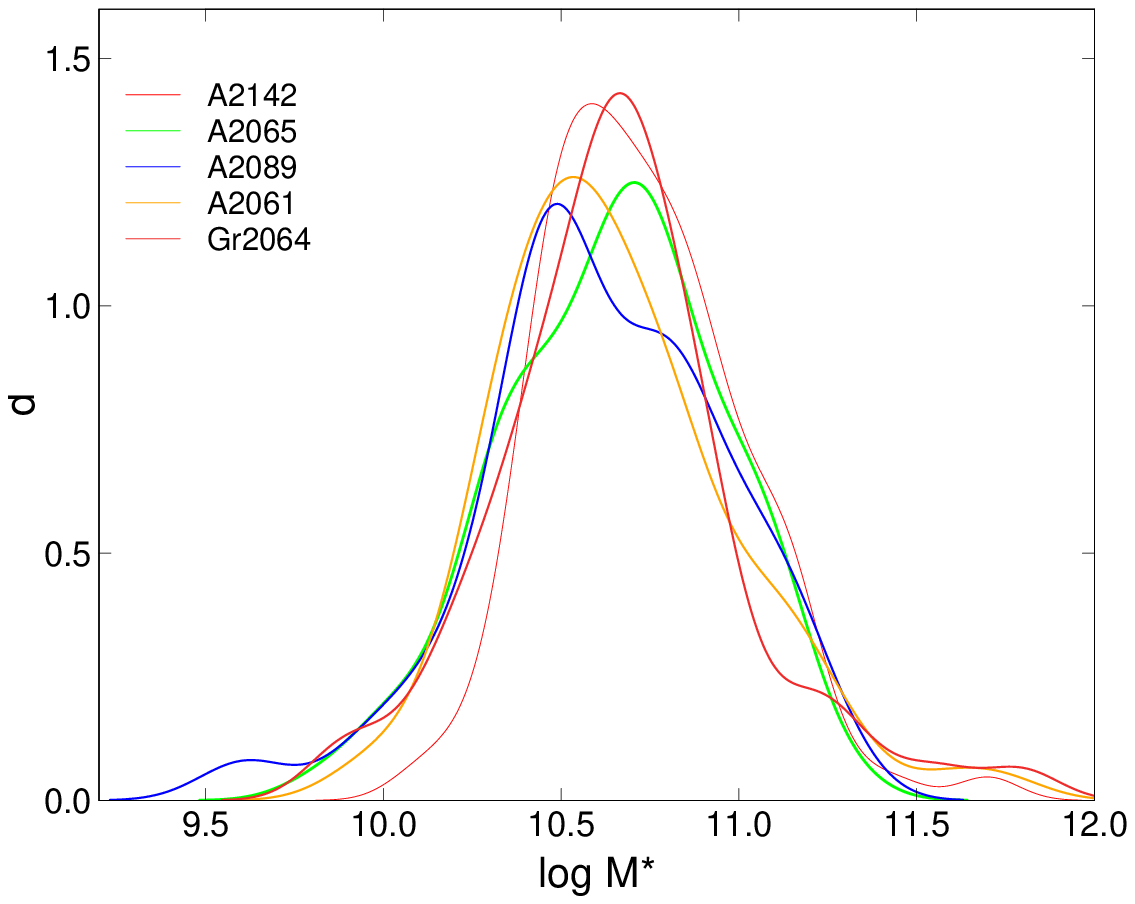}}
\caption{
Distribution of $D_n(4000)$ index of galaxies (left panel) and 
distribution of stellar masses of galaxies (right panel) in clusters A2065, A2061, A2089,
Gr2064, and A2142.
Vertical lines in the left panel 
show the limits $D_n(4000) = 1.75$, $D_n(4000) = 1.55$, and $D_n(4000) = 1.35$.
}
\label{fig:cl4dn4}
\end{figure}

\begin{figure}[ht]
\centering
\resizebox{0.23\textwidth}{!}{\includegraphics[angle=0]{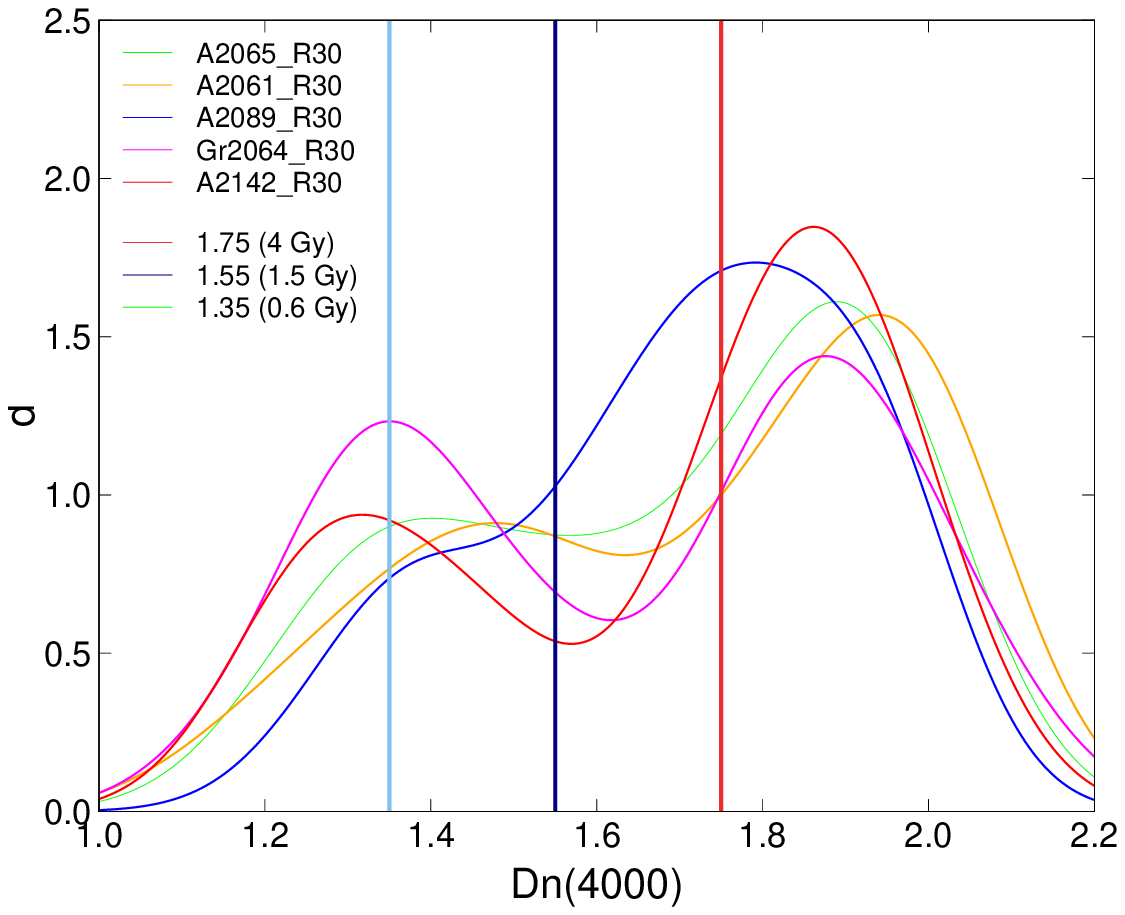}}
\resizebox{0.23\textwidth}{!}{\includegraphics[angle=0]{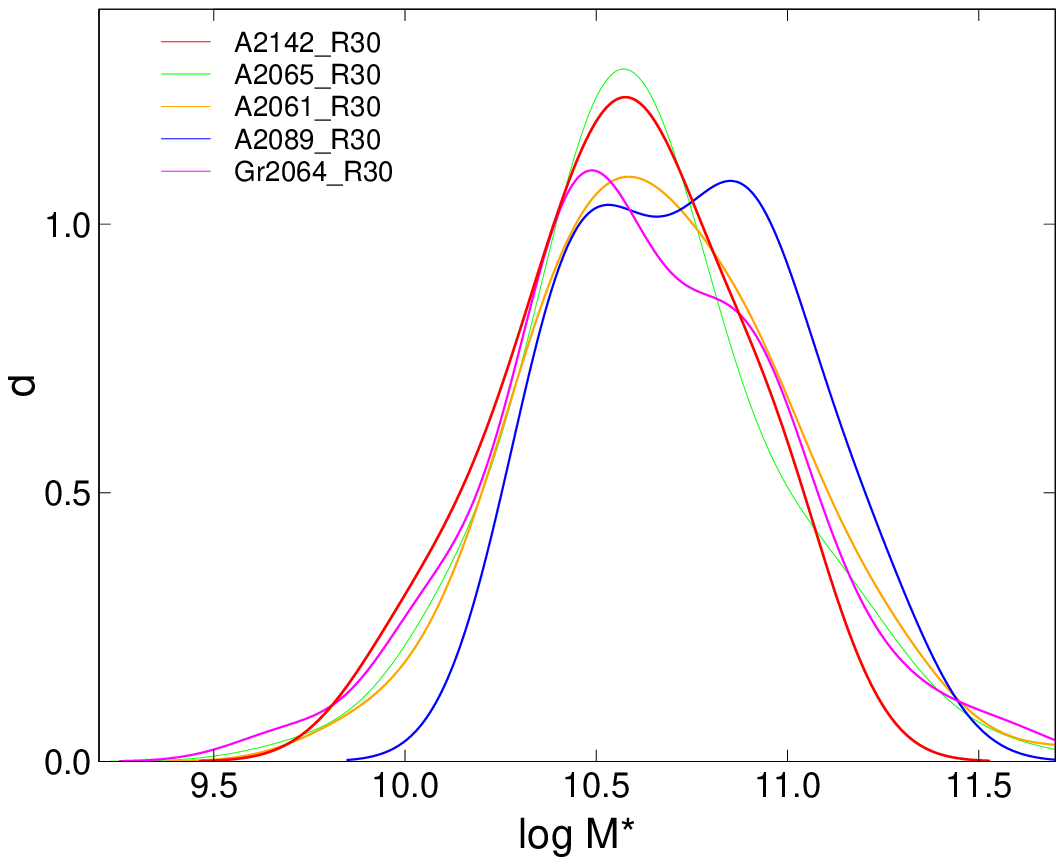}}
\caption{
Distribution of $D_n(4000)$ index of galaxies (left panel) and 
distribution of stellar masses of galaxies (right panel) in the environment of clusters A2065, A2061, A2089,
Gr2064, and A2142.
Vertical lines in the left panel 
show the limits $D_n(4000) = 1.75$, $D_n(4000) = 1.55$, and $D_n(4000) = 1.35$.
}
\label{fig:cl4envdn4}
\end{figure}

Figure~\ref{fig:cl4dn4} shows the distributions of the
$D_n(4000)$ index (left panel) and stellar masses $\log M^{\mathrm{*}}$ (right panel)
in clusters A2065, A2061, A2089, Gr2064, and A2142. In Fig.~\ref{fig:cl4envdn4} we present 
these distributions for galaxies in the spheres of influence of these clusters.
In Table~\ref{tab:galpopclr30} we give median values of 
$D_n(4000)$ index and stellar masses $\log M^{\mathrm{*}}$ of the galaxies in the clusters and
in their regions of influence, and the fractions of galaxies 
with $D_n(4000) \geq 1.75$ (VO galaxies, $f_{1.75}$), 
$1.35 < D_n(4000) \leq 1.75$ (RQ and RSF galaxies, and  BSF galaxies 
in this $D_n(4000)$ index interval, $f_{1.55}$), and
$D_n(4000) \leq 1.35$ (galaxies with the youngest stellar populations, $f_{1.35}$).
In the $D_n(4000)$ interval $1.35 < D_n(4000) \leq 1.75$ we do not separate
RQ and RSF galaxies because they have been discussed briefly in Sect.~\ref{sect:results} 
in the subsections about each individual cluster. 

\begin{table}[ht]
\caption{Median values of the $D_n(4000)$ index and stellar masses
(${\mathrm{log}} M^{\mathrm{*}}$), and the fractions of galaxies with
$D_n(4000) \geq 1.75$ ($f_{1.75}$), $1.35 < D_n(4000) \leq 1.75$ ($f_{1.55}$),
and $D_n(4000) \leq 1.35$ ($f_{1.35}$)
for galaxies in the CB and SCl~A2142 clusters and in their spheres of influence, $R_{30}$.}
\begin{tabular}{lrrrrrr} 
\hline\hline  
(1)&(2)&(3)&(4)&(5)&(6)&(7)\\      
\hline 
 $ID$ & $N_{\mathrm{gal_v}}$ & $Dn_{\mathrm{med}}$ & $f_{1.75}$ &  $f_{1.55}$ & $f_{1.35}$ 
 & ${\mathrm{log}} M^{\mathrm{*}}_{\mathrm{med}}$\\
\hline
Clusters \\
\hline
A2065           & 103 & 1.84 & 0.61 & 0.25 & 0.14 & 10.66\\
A2089           &  40 & 1.87 & 0.60 & 0.35 & 0.05 & 10.56\\
A2061           &  81 & 1.91 & 0.72 & 0.17 & 0.11 & 10.61\\
Gr2064          &  61 & 1.88 & 0.64 & 0.28 & 0.08 & 10.65\\
A2142           & 197 & 1.88 & 0.72 & 0.23 & 0.06 & 10.70\\
\hline
$R30$ \\
\hline
$A2065_{R30}$   & 210 & 1.75 & 0.50 & 0.35 & 0.15 & 10.60\\
$A2089_{R30}$   &  23 & 1.73 & 0.48 & 0.48 & 0.04 & 10.72\\
$A2061_{R30}$   & 109 & 1.80 & 0.54 & 0.34 & 0.12 & 10.64\\
$Gr2064_{R30}$  &  48 & 1.73 & 0.50 & 0.29 & 0.21 & 10.57\\
$A2142_{R30}$   &  49 & 1.80 & 0.53 & 0.27 & 0.20 & 10.59\\
\hline
\label{tab:galpopclr30}  
\end{tabular}\\
\tablefoot{                                                                                 
Columns are as follows:
(1): Cluster ID;
(2): Number of galaxies in a cluster (with $M_r = -19.6$~mag);
(3): Median value of the $D_n(4000)$ index for galaxies in a cluster;
(4--6): Fractions of galaxies with
$D_n(4000) \geq 1.75$ ($f_{1.75}$), $1.35 < D_n(4000) \leq 1.75$ ($f_{1.55}$),
and $D_n(4000) \leq 1.35$ ($f_{1.35}$) in a cluster; 
(7): Median value of the stellar masses
(${\mathrm{log}} M^{\mathrm{*}}$) of galaxies in a cluster. 
}
\end{table}

In Fig.~\ref{fig:cl4dn4} the distributions of the $D_n(4000)$ indexes 
show that clusters A2061 and Gr2064 have galaxies with a higher $D_n(4000)$ index
(higher value of $f_{1.75}$, indicating that galaxies have older stellar populations) than
clusters A2065 and A2089. 
The KS test shows that  
these differences are significant. The galaxy populations
(according to $D_n(4000)$ index) in A2061 and in Gr2064
are statistically similar, although the cluster Gr2064 has higher fraction of galaxies 
with intermediate values of the $D_n(4000)$ index than A2061
($f_{1.55}$).
The KS test shows that galaxy populations
in A2065 and in A2089  are also statistically similar, although the fraction
of galaxies with very young stellar populations is higher in A2065. 
Cluster A2089 has the highest fraction of galaxies with
$1.35 < D_n(4000) \leq 1.75$. These are mostly red and blue 
star-forming galaxies (Fig.~\ref{fig:a2089radecpps}).
As the connectivity of the cluster A2065 is as high as $\pazocal{C} = 9$ 
(Table~\ref{tab:clfil}) and the cluster contains 
infalling substructures and groups, 
we may conclude that A2065 is a dynamically active cluster, and
this may explain the high fraction of star-forming
galaxies in it. 

We found that the galaxy populations in A2142 and in A2061 are statistically similar. 
In both clusters, the stellar populations are older than in A2065 and A2089,
although A2061 has higher fraction of galaxies with young stellar populations
with $D_n(4000) \leq 1.35$ (11\% in A2061 and 6\% in A2142). These galaxies 
lie in the infall region of A2061 (Fig.~\ref{fig:a2061radecpps}).

The KS test
shows that the stellar masses of the galaxies are statistically similar in all clusters. However, we note the lack 
of galaxies with low stellar masses in clusters A2142, A2061, and Gr2064, which also had
galaxies with older stellar populations than other clusters.  
Cluster A2089 has a larger number of low stellar mass galaxies than other
clusters. 

In the spheres of influence, the fraction of galaxies with very old stellar
populations is lower than in clusters. This is expected because 
as galaxies fall into clusters, they loose their gas and their star formation
is suppressed. 
It is interesting to note that galaxies in the environment of
A2065 and A2089 
within $R_{30}$ have lower values of the
$D_n(4000)$ index than galaxies in the spheres of influence of other
clusters. This difference is statistically significant. 
The neighbourhood of Gr2064 is different from the neighbourhood of any other
cluster in this study, with an excess of 
galaxies with $D_n(4000) < 1.55$, that is, galaxies with young
and very young stellar populations. As we showed in Fig.~\ref{fig:gr2064radecpps},
there is an infalling rich group near Gr2064 with a large number 
of star-forming galaxies, and this excess is partly due to galaxies in this 
group. We return to this in the discussion.

Figure~\ref{fig:cl4envdn4} shows that in the environment
of clusters, there is some lack of low stellar mass galaxies around A2061 and around
A2089, in comparison with the environment of Gr2064 and A2065.
However, the KS test shows that these differences are not significant.
\citet{2018A&A...620A.149E} noted that there is a large variety in the galaxy content
of individual structures falling into the cluster A2142, but on average, the galaxy content
in the sphere of influence of A2142 is similar to that around the other clusters we studied.

\subsection{Summary of cluster properties}
\label{sect:clsummary}

\begin{figure*}[ht]
\centering
\resizebox{0.86\textwidth}{!}{\includegraphics[angle=0]{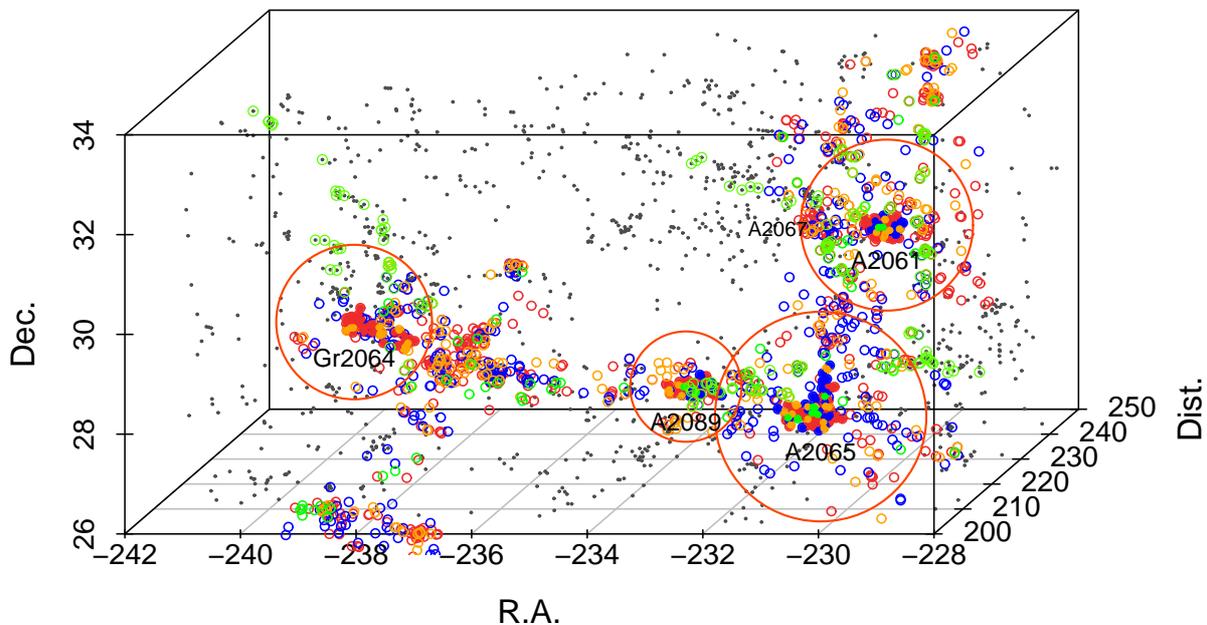}}
\caption{
3D distribution of galaxies in the CB and in the low-density region around it.
Filled symbols denote galaxies in clusters A2065, A2061, A2089, and Gr2064.
Empty circles show supercluster galaxies that are not members of these clusters. 
Grey dots denote galaxies in the low-density region with a global density $D8 < 5$.  
Blue circles denote BSF galaxies, orange circles denote RSF galaxies, turquoise
green circles mark RQ galaxies, and red circles denote VO galaxies. 
Green circles mark galaxies in long filaments with a length  larger than 5~\Mpc.
Large red circles show the circles with a radius $R_{\mathrm{30}}$,
which partly overlap because of projections in the 3D view.
}
\label{fig:cb3d}
\end{figure*}

We analysed the dynamical state, connectivity, and galaxy populations
of the richest clusters in the high-density cores of the CB supercluster.
We show the 3D distribution of galaxies in the CB region in Fig.~\ref{fig:cb3d}.
In our study we found that at a certain 
clustercentric distance around each cluster, 
there is a minimum in the distributions of
galaxies. According to the PPS diagrams, within this distance limit, 
galaxies and groups are falling into the central cluster.
We may assume that the corresponding clustercentric distance  
marks the borders of the sphere of influence for clusters.
We denoted this distance as the radius $R_{30}$ (Table~\ref{tab:clfil}). 
The spheres of influence of 
three clusters touch each other, but they do not overlap
(they overlap partly in Fig.~\ref{fig:cb3d} because of
projections).  
Interestingly, a similar scale has been found in the distribution of galaxy 
groups along filaments \citep{2014A&A...572A...8T}.
\citet{2020A&A...641A.172E} noted a similar feature around the cluster A2142 in the high-density
core of the SCl~A2142 supercluster, where it  corresponds to the density
contrast around the cluster $\Delta\rho \approx 30$. To compare the density contrast around
clusters from the CB and in SCl~A2142, we calculate corresponding
density contrasts for the CB clusters below, but first we briefly
summarise the results for the clusters.

A2065 is the richest and most massive cluster in the CB. It has the largest number of substructures,
the highest connectivity 
and the second highest number of long filaments of the CB clusters. The 
star formation activity both in the cluster and in its sphere of influence
may be triggered by the overall dynamical activity of the cluster.
Infalls into A2065 occur from many directions: 
along the line of sight, and from north and east, 
where A2061 and A2089 lie.

A2061 is the second richest cluster in the CB. Its mass is close to the mass of Gr2064.
A2061 forms a close pair with A2067, and these clusters have 
probably already started to merge. It has the largest number of long filaments that reach out 
of its sphere of influence.  
Relatively old stellar populations both in the cluster and in its
sphere of influence suggest that it already passed its most active period
of star formation or that the transition of galaxies is now ongoing, as indicated by the
high fraction of (red) star-forming galaxies near it.

The cluster A2089 is the poorest and has the lowest mass of the clusters we studied.
It also has the lowest connectivity value and the smallest $R_{30}$. 
There is only one long filament near it that 
points towards the cluster A2065. A2089 itself is elongated in the direction of this filament.
A2089 is dominated by galaxies with old stellar populations. We may assume that
this cluster has already passed the active formation stage. 

Cluster Gr2064 is the richest cluster in the high-density peak in an
another part of the CB. Its mass is similar to the mass of A2061.
It has three long filaments, one of which connects A2064 with the other rich groups in CB. 
Galaxy populations of galaxies in the sphere of influence
of Gr2064 differ from those around any other rich cluster in the CB.
This cluster has an infalling group with the highest fraction of star-forming galaxies
of the infalling groups near the CB clusters. There are also RSF galaxies in the outskirts of
GR2064. 

In all clusters at $R_{\mathrm{cl}}$ there is a small minimum in the galaxy
distribution followed by a local maximum. This is similar to the splashback feature 
detected in cluster profiles 
\citep[][and references therein]{2015ApJ...810...36M, 2020arXiv201005920B}.
However, because we lack information about galaxy orbits, we call this 
the radius of the (main) cluster $R_{\mathrm{cl}}$, where substructures and 
groups or clusters near clusters are falling into the cluster. 
We defined the radius $R_{\mathrm{cl}}$ 
for each cluster empirically as the radius at which the minimum in the galaxy
distribution occurs.
The values of $R_{\mathrm{cl}}$ lie between the virial radii of clusters and
their maximum size in the sky, given in Table~\ref{tab:cl}. They depend on the
structure and merging history of clusters.
Some galaxies
here may actually be splashback galaxies, but we need to study them in more detail
to test this. This could be a subject for future studies.

Several studies have found that dynamically active
clusters have higher fraction of star-forming galaxies than relaxed clusters
\citep[see, for example, ][]{2010A&A...522A..92E, 2014ApJ...783..136C, 
2017A&A...607A.131D, 2020MNRAS.497..466S, 2021arXiv210206612S}.
Star formation of 
galaxies in these clusters may be triggered by infall of galaxies and groups.
In our study we found that clusters A2061 and A2065 have the largest
number of infalling structures and long filaments near them,
with connectivities $\pazocal{C} = 8$ and $\pazocal{C} = 9$, respectively, 
but their galaxy populations are different. Cluster A2061 has galaxies with older
stellar populations than A2065.
This suggests that the formation history of these clusters has been different.

Interestingly, the galaxy populations in A2142 and in A2061 are statistically similar. 
In both clusters the stellar populations are older than in A2065 and A2089.
\citet{2018A&A...610A..82E} showed that on average, galaxies in
A2142 are redder and have lower star formation rates than other rich clusters
from the SDSS at the same distances. Our finding that the
galaxy populations in A2142 and A2061 are similar shows that
large sets of data need to be compared with caution: individual differences and possible 
large variations in the properties of the objects, in our case, 
the richest galaxy clusters in superclusters, might be washed out. Large variations in the galaxy
populations of individual rich clusters in the same supercluster have been found before, for example, in the superclusters of the Sloan Great Wall and in the
Coma supercluster \citep{2010A&A...522A..92E, 2020MNRAS.497..466S}.
Moreover, variations in star formation properties of galaxies
have been found in infalling systems near clusters A2142 and A963
\citep{2010A&A...522A..92E, 2017A&A...607A.131D,
2018A&A...610A..82E, 2020A&A...638A.126D}. 
This is probably related to the 
different history of these galaxies \citep{2013MNRAS.430.3017B, 2020MNRAS.497..466S}.
These are interesting topics for the future studies
that search for variations in cluster properties based on simulations.

We also found a certain cosmic concordance of the galaxy properties
in clusters and in their spheres of influence in the main part of the CB. 
Clusters with a higher fraction of
star-forming galaxies also have a higher fraction of star-forming galaxies 
in their neighbourhood. However, the environment of Gr2064 
is different, with an infalling group with a very high fraction of star-forming
galaxies. This is again evidence of large variations in galaxy populations
of individual rich clusters and their spheres of influence,
which shows that the clusters in the CB have different merging histories.

\begin{table}[ht]
\caption{Characteristic radii $R_{\mathrm{vir}}$, $R_{\mathrm{cl}}$ , and
$R_{\mathrm{30}}$, and connectivity $\pazocal{C}$ of clusters.}
\begin{tabular}{lrrrrrr} 
\hline\hline  
(1)&(2)&(3)&(4)&(5)&(6)&(7)\\      
\hline 
No. & ID& $R_{\mathrm{vir}}$ & $R_{\mathrm{cl}}$ & $R_{\mathrm{30}}$ &$\pazocal{C}$ & $N_{\mathrm{fil}}$\\
\hline                                                    
 1 &A2065  & 0.7 & 2.5& 6 & 9& 4\\
 2 &A2061  & 0.5 & 1.4& 4 & 8& 5\\
 3 &A2089  & 0.6 & 1.3& 3 & 2& 1\\
 4 &Gr2064 & 0.7 & 1.5& 4 & 6& 3\\
\hline                                        
\label{tab:clfil}  
\end{tabular}\\
\tablefoot{                                                                                 
Columns are as follows:
(1-2): Order number and Abell ID of the cluster;
(3): Cluster virial radius $R_{\mathrm{vir}}$ (in \Mpc);
(4): Cluster radius $R_{\mathrm{cl}}$ (in \Mpc);
(5): Radius of the sphere of influence, $R_{\mathrm{30}}$ (in \Mpc);
(6): Total connectivity $\pazocal{C}$ of the cluster (the number of filaments,
groups, and substructures);
(7): Number of long filaments with a length greater than $5$~\Mpc\ near the cluster,
$N_{\mathrm{fil}}$. 
}
\end{table}

When we count all long filaments that extend from the CB supercluster
to the low-density region around it \citep[supercluster cocoon, see ][for details and references]
{2020A&A...641A.172E}, then we find that the connectivity of the CB is 
$\pazocal{C} = 7$ (here we did not count the long filaments between clusters
inside the supercluster). Thus the connectivity of the CB is comparable with that
of SCl~A2142 with $\pazocal{C} = 7$. 

\section{Discussion}
\label{sect:discussion} 

\subsection{Spheres of influence $R_{30}$ and the density contrast $\Delta\rho$}
\label{sect:sphden}  

\begin{figure}[ht]
\centering
\resizebox{0.47\textwidth}{!}{\includegraphics[angle=0]{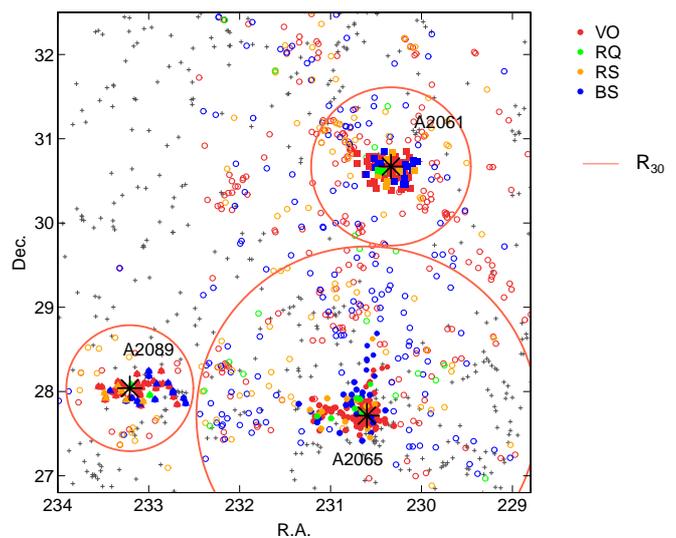}}
\caption{
Distribution of galaxies in the main part of the CB in the plane of the sky. 
 Filled symbols denote galaxies in clusters A2065, A2061,
and A2089, and empty circles show other galaxies in the supercluster. Grey crosses
denote galaxies in the low-density region with $D8 < 5$ (partly overlapping with the supercluster
region due to projections).  
Large red circles show the circles with radius of $R_{\mathrm{30}}$ (Table~\ref{tab:clfil}).
}
\label{fig:cbwradecr30}
\end{figure}

In Fig.~\ref{fig:cbwradecr30} we plot the sky distribution of galaxies 
in the main part of the CB with $R.A. < 234$~degrees and plot the circles around each
cluster showing the spheres of influence of the corresponding cluster.  
This figure shows that spheres of influence
do not overlap. This is expected from the evolution of protoclusters, 
where clusters grow by accretion of galaxies from surrounding region
\citep{2013ApJ...779..127C, 2016A&ARv..24...14O}. Each cluster is surrounded 
by a region from which galaxies fall towards the central cluster, as also
seen in the PPS diagrams above. When we compare this figure with Fig.~\ref{fig:cbradec}, 
we see that the borders of the regions of influence approximately correspond
to the minima in the luminosity-density field in the CB (we discuss this in
more detail in another study).

\begin{figure}[ht]
\centering
\resizebox{0.47\textwidth}{!}{\includegraphics[angle=0]{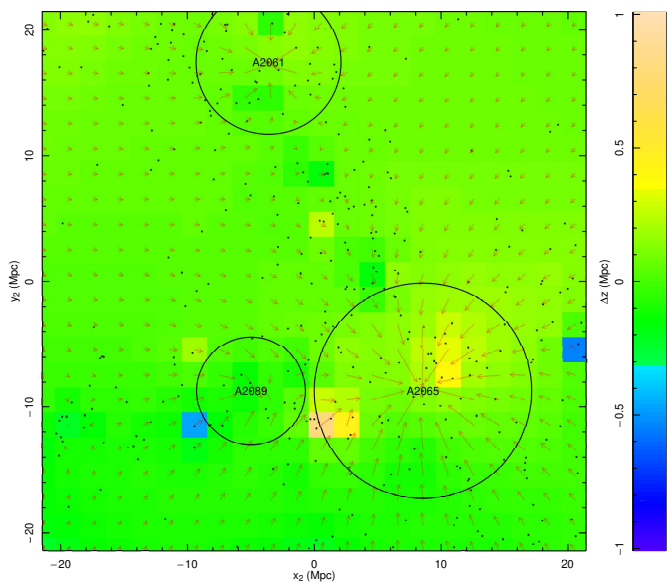}}
\caption{
Acceleration field of the CB supercluster. 
Here  coordinates $(x,y)$ are rectangular coordinates in the plane 
containing centres of all three main clusters. Colours show the values of
the difference in $z$ coordinate direction, and arrows show how far galaxy groups 
in the CB may move during $2$ Gyr. 
}
\label{fig:accfield}
\end{figure}

In Fig.~\ref{fig:accfield} we show the acceleration field of the CB supercluster. 
Here arrows show how far galaxy groups in the CB may move during $2$ Gyr. 
Coordinates $(x,y)$ are rectangular coordinates in the plane containing 
the centres of all three main clusters. 
This is quite different from the plane of the sky, thus the sphere of influence
around A2061, seen at a different angle, appears to be smaller than in Fig.~\ref{fig:cbwradecr30}.
The spheres of influence around A2065 and A2089 touch, as in Fig.~\ref{fig:cbwradecr30}.
In the acceleration field figure (Fig.~\ref{fig:accfield}), 
each cluster acts as a small attractor.
The centre of the gravitational potential lies between clusters.
The separation between the clusters, where the acceleration is
lowest, is seen in the field between clusters. 
In calculations of the acceleration field, all four richest
clusters in the CB have been taken into account. We do not show the cluster Gr2064 
in Fig.~\ref{fig:accfield}  because it is
difficult to project it onto the plane that is determined by
the three other richest clusters in the CB without distortions.

We note that \citet{2020A&A...641A.172E} showed that
the A2142 supercluster has a high-density core with a radius of
$\approx 5$~\Mpc, with a clear
minimum in the galaxy distribution. This minimum separates it from the 
surrounding main body of the supercluster.
\citet{2020A&A...641A.172E} found that the minimum corresponds 
to the density contrast $\Delta\rho \approx 30$.
It is interesting to analyse the density contrast around rich clusters in the CB and
to compare it with that of the SCl~A2142.

For this purpose, we compared the mass distribution around clusters
with the predictions of the spherical collapse model.
In our analysis we used the density contrast $\Delta\rho$ within spheres around
clusters with increasing radius.
To calculate the density contrast, we used the distribution of 
cluster and group masses calculated as  described
in Sect.~\ref{sect:sph}.  From the cumulative mass distribution, 
we calculated the corresponding density distributions.

\begin{table*}[ht]
\caption{Radii and masses of centre clusters within their spheres of influence
at characteristic density contrasts $\Delta\rho = 30$, $13.1$, and $8.7$.}
\label{tab:clmasses}  
\begin{tabular}{rrrrrrrrrrr} 
\hline\hline  
(1)&(2)&(3)&(4)&(5)& (6)&(7)&(8) & (9)&(10)&(11)\\      
\hline 
No. & ID & $R_{\mathrm{vir}}$&  $M_{\mathrm{dyn}}$ &  $M_{\mathrm{30}}$& $R_{\mathrm{30}}$
& $R_{\mathrm{turn}}$&  $M_{\mathrm{turn}}$ 
& $R_{\mathrm{FC}}$&  $M_{\mathrm{FC}}$& $D_{\mathrm{2065}}$\\
\hline                                                    
 1 &A2065 & 0.7& 1.5 & 2.6 & 6  & 9.5 & 3.7 & 11.9 & 4.3 & 0\\
 2 &A2061 & 0.5& 0.4 & 0.9 & 4  & 6.4 & 1.2 &  8.6 & 1.8 & 11.0\\
 3 &A2089 & 0.6& 0.2 & 0.5 & 3  & 5.6 & 0.8 &  6.9 & 0.9 & 8.7\\
 4 &Gr2064& 0.7& 0.3 & 0.7 & 4  & 4.9 & 0.7 &  8.5 & 1.6 & 29.4\\
\hline                                        
\end{tabular}\\
\tablefoot{                                                                                 
Columns are as follows:
(1): Order number of the cluster;
(2): ID  of the cluster;
(3): Cluster virial radius (in $h^{-1}$ Mpc);
(4): Dynamical mass of the cluster, $M_{\mathrm{dyn}}$
(in $10^{15}h^{-1}M_\odot$);
(5): Radius $R_{\mathrm{30}}$;
(6): Mass $M_{\mathrm{30}}$ 
(in $10^{15}h^{-1}M_\odot$), embedded in a sphere around the cluster with a radius of $R_{\mathrm{30}}$;
(7): Radius $R_{\mathrm{turn}}$ (radius of the turnaround region around the cluster);
(8): Mass $M_{\mathrm{turn}}$, embedded in a sphere around the cluster with a radius of $R_{\mathrm{turn}}$;
(9): Radius $R_{\mathrm{FC}}$ (radius of the future collapse region around the cluster);
(10): Mass $M_{\mathrm{FC}}$, embedded in a sphere around the cluster with a radius of $R_{\mathrm{FC}}$;
(11): Distance from the centre of A2065, $D_{\mathrm{2065}}$ (in $h^{-1}$ Mpc).
}
\end{table*}

\begin{figure*}[ht]
\centering
\resizebox{0.40\textwidth}{!}{\includegraphics[angle=0]{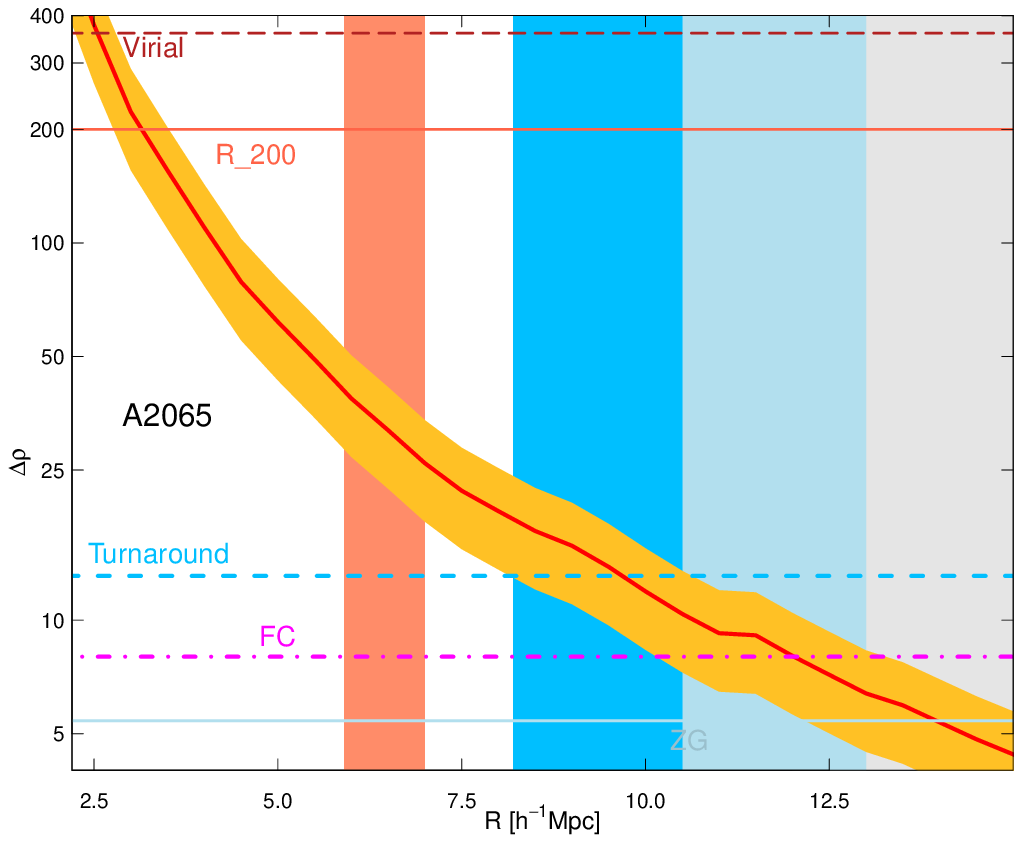}}
\resizebox{0.40\textwidth}{!}{\includegraphics[angle=0]{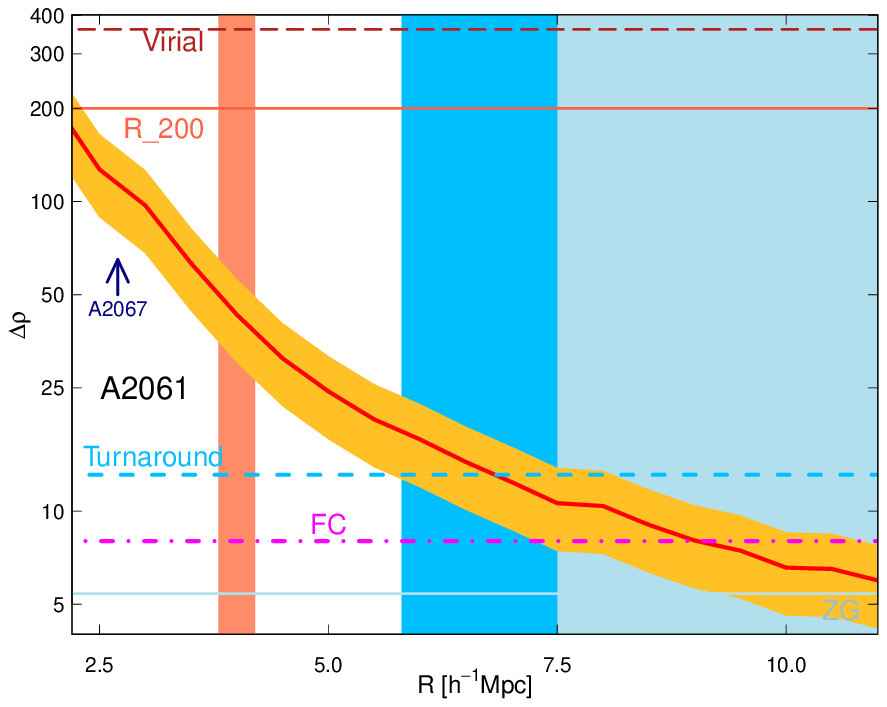}}
\caption{
Density contrast $\Delta\rho = \rho/\rho_{\mathrm{m}}$ vs.
clustercentric distance  for A2065 (left panel) and for A2061 (rich panel).
The density contrast for the centre cluster is plotted with the red line. 
The golden area shows the error corridor
from the mass errors. Characteristic density contrasts are
denoted as follows: 
$\Delta\rho = 360$ (virial),
$\Delta\rho = 200$ ($r_{200}$), $\Delta\rho = 13.1$ (turnaround, dashed blue line),
$\Delta\rho = 8.73$ (FC, dash-dotted violet line), 
and $\Delta\rho = 5.41$
(ZG, solid light blue line).
Blue, light blue, and grey areas mark the borders
of the turnaround, FC, and ZG regions. 
The red area marks $R_{30}$.
The arrow in the right panel marks the average
distance of cluster A2067 from the centre of A2061.
}
\label{fig:dcrho20652061}
\end{figure*}

\begin{figure*}[ht]
\centering
\resizebox{0.40\textwidth}{!}{\includegraphics[angle=0]{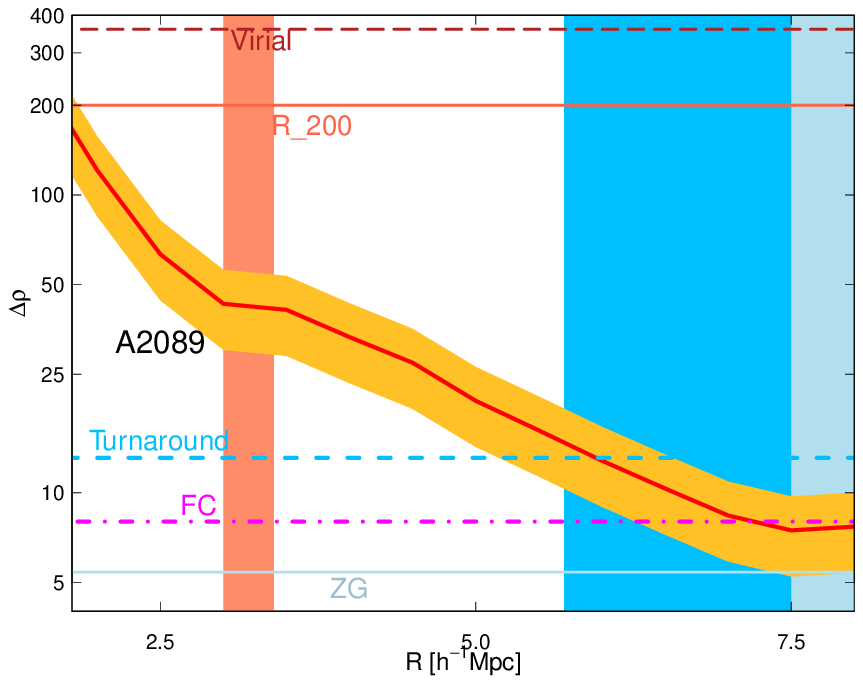}}
\resizebox{0.40\textwidth}{!}{\includegraphics[angle=0]{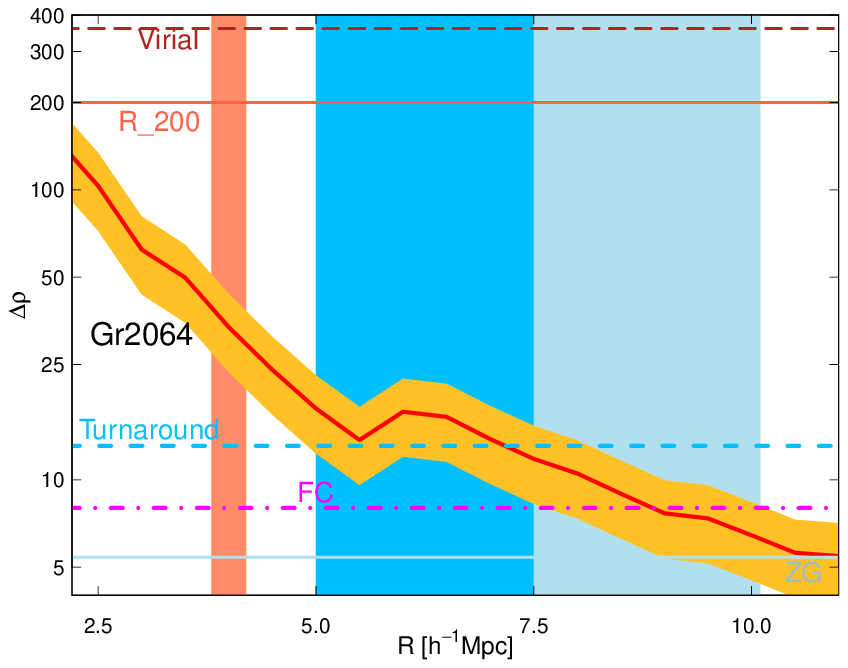}}
\caption{
Density contrast $\Delta\rho = \rho/\rho_{\mathrm{m}}$ vs.
clustercentric distance  for cluster A2089 (left panel)
and for Gr2064 (left panel). 
Notations are as in Fig.~\ref{fig:dcrho20652061}.
}
\label{fig:dcrho2089}
\end{figure*}

In Figs.~\ref{fig:dcrho20652061} and \ref{fig:dcrho2089} we plot the
density contrast $\Delta\rho$ with respect to the clustercentric distance
for A2065, A2061, A2089, and for Gr2064.
In the right panel of Fig.~\ref{fig:dcrho20652061} we mark the location
of A2067 from the centre of A2061.
In these figures vertical red areas  correspond 
to the borders of the spheres of influence, $R_{\mathrm{30}}$.
It is interesting to note that they 
all approximately correspond to the density contrast $\Delta\rho\simeq 30$
(or slightly higher).
The same density contrast was found at the borders of the high-density core
of the SCl~A2142 \citep{2020A&A...641A.172E}.
Vertical blue, light blue, and grey areas in Figs.~\ref{fig:dcrho20652061} and 
\ref{fig:dcrho2089} mark borders 
of the turnaround, future collapse, and zero-gravity regions.
The width of these areas is determined by the width of the mass error corridor.
These regions were introduced in Sect.~\ref{sect:sph}. The turnaround
radius corresponds to the radius of a spherical shell around a cluster 
that is at the turnaround. Within this radius, the galaxy systems around clusters
(between $R_{\mathrm{30}}$ and $R_{\mathrm{turn}}$)
are collapsing, that is, they fall into the clusters. 
Galaxy systems within the future collapse region will collapse  in the future.

The density contrast at the turnaround in the local Universe is 
$\Delta\rho = 13.1$. 
The density contrast $\Delta\rho$ at the borders of the spheres
of influence, $\Delta\rho \simeq 30$, shows that
the structures within 
$R_{30}$ around clusters have already passed the turnaround some time ago,
and they continue to contract. 
This might be a universal density contrast that characterises the evolution of clusters.
To understand at which redshift the collapse may have begun if the current density contrast
is $\Delta\rho\simeq 30$, we analysed the evolution of the density
perturbation and calculated the current density contrast
for cases in which the
turnaround occurs at redshifts in the range of $z = 0.6- 0.2$. 
In our calculations we used formulae derived by \citet{2010JCAP...10..028L}.
We show the results for two values of $\Omega_\mathrm{m}$, $\Omega_\mathrm{m} = 0.27$,
as used in this paper, and $\Omega_\mathrm{m} = 0.4$. The reason for this is that
according to current observational data, 
the value of $\Omega_\mathrm{m}$ lies in a certain range around $0.3$ 
\citep{2020arXiv200715632H}.
For $\Omega_\mathrm{m} = 0.3$ the turnaround density contrast value can be found, for example, in
\citet{2015A&A...581A.135G} and in \citet{2002MNRAS.337.1417G}.

\begin{figure}[ht]
\centering
\resizebox{0.44\textwidth}{!}{\includegraphics[angle=0]{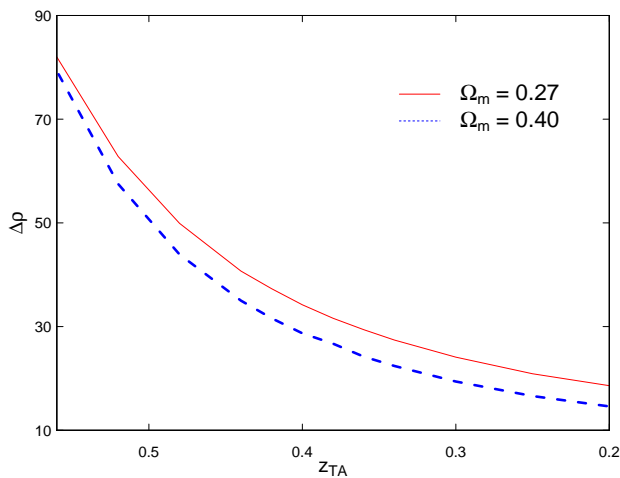}}
\caption{
Current density contrast $\Delta\rho$  for $\Omega_\mathrm{m} = 0.27$ (solid red line)
and for $\Omega_\mathrm{m} = 0.40$ (dashed blue line)
if the turnaround occurred at redshifts in the interval $z = 0.56 - 0.2$.   
}
\label{fig:rhoz}
\end{figure}

We  show the results of these
calculations in Fig.~\ref{fig:rhoz}. This figure shows the density contrast $\Delta\rho$ at present,
if the turnaround occurred at higher redshifts.   
Structures with a current value $\Delta\rho\simeq 30$
passed the turnaround at redshifts  $z \approx 0.37$ for $\Omega_\mathrm{m} = 0.27$,
and at redshifts  $z \approx 0.41$ for $\Omega_\mathrm{m} = 0.40$.
The difference between the two $\Omega_\mathrm{m}$ values is very small, as
was also concluded in \citet{2015A&A...581A.135G}.
We may therefore assume that structures around clusters within the spheres of influence, 
$R_{\mathrm{30}}$,
passed the turnaround in the redshift interval $z \approx 0.3 - 0.4$.
Galaxies closer to the clusters than this radius are falling into clusters,
and this is probably the reason for the minimum in the galaxy distribution at $R_{\mathrm{30}}$.
Protocluster
simulations show that the clusters themselves probably already obtained their
current sizes at redshifts $z > 0.4$ \citep{2013ApJ...779..127C}.
This agrees with the estimate about the formation of cluster A2142 by 
\citet{2018A&A...610A..82E}.

\citet{2020A&A...641A.172E} showed that the galaxy groups within the 
high-density core of A2142 are all falling into
cluster A2142 \citep[see][for details]{2020A&A...641A.172E}.
We showed the same for the richest clusters in the CB. 
Therefore each cluster  acts as a small attractor 
for galaxies within their sphere of influence with $R_{\mathrm{30}}$. 
The radii of the spheres of influence, $R_{\mathrm{30}}$, are proportional
to the mass of the cluster and also to the mass inside the spheres of influence, but the scatter
is large (Table~\ref{tab:clmasses}). Tables~\ref{tab:galpopclr30} and \ref{tab:clfil}
show that galaxy content and connectivity  of clusters are not directly related
to their masses or to radii  $R_{\mathrm{30}}$. In other words, 
in the high-density cores of the CB, the richest clusters and their close environment
show large variations in their dynamical properties, in addition 
to the variations in their galaxy content and connectivity. 

\subsection{Evolution of the CB}
\label{sect:cbevol}  

Next we discuss possible future evolution scenarios of the richest clusters in the CB 
and for the whole supercluster. For this we compare the density contrasts
at various clustercentric distances in Figures~\ref{fig:dcrho20652061} and \ref{fig:dcrho2089}
and in Table~\ref{tab:clmasses}
with the characteristic radii and density contrasts from the spherical collapse model.

{\bf Distancing scenario}. Figures~\ref{fig:dcrho20652061} and 
\ref{fig:dcrho2089} and Table~\ref{tab:clmasses} 
demonstrate that the turnaround radius $R_{\mathrm{turn}}$ 
for each cluster is larger than the radius
$R_{\mathrm{30}}$. This means that 
each cluster with its surrounding region has already passed turnaround.

One scenario for the future of the CB,
which can be called the {\it \textup{distancing scenario,}} therefore is that each of  clusters 
will shrink into a separate system. These systems will  move away from each other as the Universe 
expands.
Unfortunately, 
as we lack distances for galaxies that 
are independent from Hubble expansion and we therefore also lack their peculiar velocities, 
we cannot directly determine whether this scenario really occurs. 
However, cluster 
Gr2064 lies far away from the clusters in the main part of the CB.
Its distance from the nearest cluster in the main part of the CB, A2089, is $\approx 20$~\Mpc\
(Table~\ref{tab:clmasses}).
This value exceeds the values of the turnaround and future collapse for all clusters.
For this cluster, the distancing scenario therefore probably correctly predicts
its future. 

This assumption is supported by the fact that this cluster 
and the whole part of the CB surrounding it is only weakly
connected with another part of the CB by a bridge of galaxies in which the global density barely exceeds the global density limit
for defining superclusters, $D8 \simeq 5$ (Sect.~\ref{sect:gr}, and Figs.~\ref{fig:cbradec}
and \ref{fig:cb3d}). 
Cluster Gr2064 is projected onto the supercluster
SCl~A2142 in the sky distribution. These two structures, the Gr2064 part of the CB supercluster
and SCl~A2142, are connected by a filament 
\citep[see Fig.~\ref{fig:gr2064radecpps} and the figures in][]{2020A&A...641A.172E}. 
This poses the question whether the Gr2064 region might be
a part of SCl~A2142. However, as the analysis in \citet{2020A&A...641A.172E} showed,
there is a density minimum in this filament that is characteristic
of supercluster cocoon borders, and the Gr2064 region lies outside of the  
SCl~A2142 cocoon. It is interesting to note that the Gr2064 region resembles the
small Arrowhead supercluster that is
embedded between the Laniakea, the Perseus-Pisces, and the Coma superclusters
in the local Universe \citep{2015ApJ...812...17P}, and has approximately the same mass
as the turnaround region around Gr2064 (Table~\ref{tab:clmasses}).
According to the distancing scenario, the Gr2064 region may in the future become a system similar to the Arrowhead
supercluster.

{\bf The total collapse scenario}. 
For clusters in the main part of the CB, the distances between A2065 and other clusters 
given in
Table~\ref{tab:clmasses} are smaller than the radius of the future collapse region
around A2065. A2089 lies within the turnaround region of A2065. 
The distance between the centres of A2065 and A2061 is approximately  $9$~\Mpc, 
which is within the future collapse radius of A2065.

\begin{figure*}[ht]
\centering
\resizebox{0.44\textwidth}{!}{\includegraphics[angle=0]{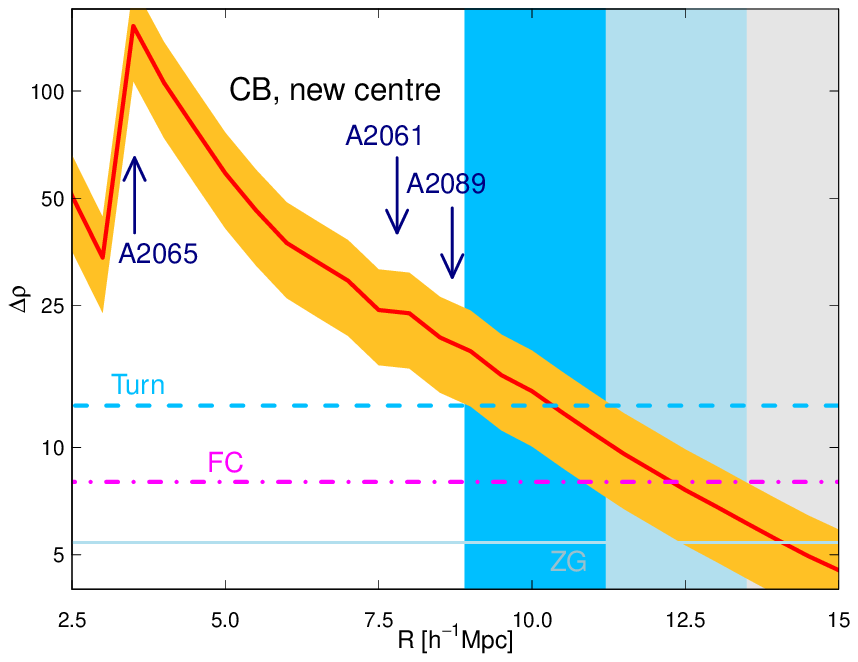}}
\resizebox{0.52\textwidth}{!}{\includegraphics[angle=0]{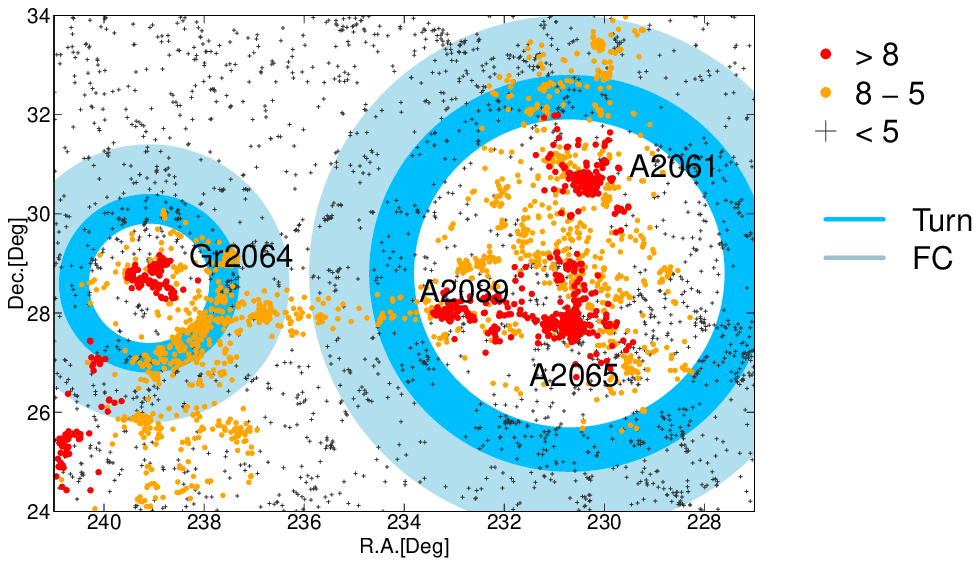}}
\caption{
Left panel: Density contrast $\Delta\rho = \rho/\rho_{\mathrm{m}}$ vs.
clustercentric distance $D_c$ for the possible new centre of the main part of the CB.
Notations are as in Fig.~\ref{fig:dcrho20652061}. Arrows mark the 
distances of clusters A2065, A2061, and A2089 at which they join the possible
new collapsing system.
Right panel: Sky distribution of galaxies in the CB. Filled circles mark turnaround 
(blue) and future collapse (light blue) zones in the supercluster,
as shown in the left panel.
Colours are as in Fig.~\ref{fig:cbradec}.
}
\label{fig:dcrhomean3}
\end{figure*}

This suggests another scenario for the future evolution of the CB, which 
can be called \textup{} the total collapse scenario. In this scenario, all three clusters will merge into one massive system. As a result, this scenario
describes the formation of 
one of the largest bound systems in the local Universe.
Cluster A2065, which  has the highest mass of the rich clusters of the CB,
will be closest to the centre of this new system.
To illustrate {\it \textup{the total collapse scenario,}} we calculated the density
contrast versus distance from the possible new centre, 
shown in the left panel of Fig.~\ref{fig:dcrhomean3}.
As in previous figures, we mark the regions of turnaround, future collapse, and zero gravity.
In the right panel of Fig.~\ref{fig:dcrhomean3} we show the sky distribution 
of galaxies in the CB, with circles corresponding to the turnaround and future collapse regions.
In the main part of the CB, they are centred on the possible new centre of the CB,
and in the Gr2064 part, they are centred on cluster Gr2064.
This division is based on the previous results of the distancing scenario,
which showed that most likely, the part of the CB with GR2064 will separate
from the main part of the CB in the future.

In this figure we mark  the distances at which the clusters join this possible future system.
The three richest clusters in the CB are within the
turnaround radius of the possible new centre.
This supports the scenario according to which the main part of the CB is already collapsing.
Its outer parts will collapse in the future. 
This figure shows that the mass of the possible new system at the turnaround radius,
$R_{\mathrm{turn}} \approx 10$~\Mpc, is $M_{\mathrm{turn}} \approx 4.5\times~10^{15}M_\odot$.
The radius and mass of the future collapse region is
$R_{\mathrm{FC}} \approx 12.5$~\Mpc\  and $M_{\mathrm{FC}} \approx 4.7\times~10^{15}M_\odot$.
Comparison with the sizes and masses of turnaround and future collapse regions around
individual clusters in Table~\ref{tab:clmasses} shows that if the centre of the
possible new system is at the most massive cluster in the CB, cluster A2065,
the total mass in the system that will collapse in the future is slightly lower
($M_{\mathrm{FC}} \approx 4.3\times~10^{15}M_\odot$), but all three clusters
lie within the radius $R_{\mathrm{FC}}$ from A2065. The lower mass comes from the fact
that in this case, a
larger part of the low-density region that surrounds the supercluster is included in the
collapsing system. With the new centre, the collapsing region mostly includes the main part of the
CB. The zero-gravity zone surrounds the whole main part of the CB. 
This is similar to what \citet{2020A&A...641A.172E} found for the 
main body of the supercluster SCl~A2142.
The total collapse scenario suggests that  
different parts of the CB will separate in the future. 

The right panel of Fig.~\ref{fig:dcrhomean3} shows 
that the turnaround region around cluster Gr2064 includes a rich group that is infalling
on Gr2064 (see Sect.~\ref{sect:gr2064}). Interestingly, above we found that
differently from the other clusters we studied, the infall zone of Gr2064 contains
a large number of star-forming galaxies (Fig.~\ref{fig:gr2064radecpps}, lower right panel).
The star formation of galaxies in this group might be enhanced due to the joint
effect of infall and collapse of this region.
Similarly, there is an excess of star-forming galaxies
in the turnaround region of supercluster SCl~A2142 (\citet{2018A&A...620A.149E})
that supports this assumption.

The total collapse scenario is similar to what \citet{2009MNRAS.399...97A}
described for the future evolution of superclusters. According to \citet{2009MNRAS.399...97A}, individual clusters and groups in superclusters will merge to form massive
systems of low multiplicity. The shapes of future superclusters will be more
spherical than the current shapes of superclusers, which are mostly 
elongated \citep{2011A&A...532A...5E}.

Our suggestions for the future evolution of the CB depend on how good our mass estimates for the supercluster are.
Our mass estimates for the Corona Borealis supercluster high-density cores
agree well with the lower limits of the mass estimates by \citet{2014MNRAS.441.1601P}.
\citet{2014MNRAS.441.1601P} reported that the A2065 region will form a bound system
in the future 
if there is enough mass between the clusters.
In Sect.~\ref{sect:mass} we showed that the mass of the supercluster may be underestimated,
partly due to the faint, unobserved galaxies. 
Therefore we may conclude that at least we probably do not overestimate
the supercluster mass, and this adds weight to our conclusions about the total collapse scenario.

The collapsing high-density cores have also been studied in other superclusters:
in the Shapley supercluster, in superclusters of the Sloan Great Wall,
in the Perseus-Pisces supercluster, and in the Hercules supercluster
\citep{2000AJ....120..523R, 2002AJ....124.1266R, 2015A&A...575L..14C, 2015A&A...577A.144T,
2016A&A...595A..70E}. High-density collapsing cores of superclusters
may inhabit several (merging) clusters, like in the Shapley supercluster.
Rich superclusters may have several cores that may collapse and form 
separate systems in the future, like in the Sloan Great Wall superclusters
\citep{2016A&A...595A..70E}. The CB supercluster is interesting among
other rich superclusters as it embeds rich clusters at various stages of
evolution, from single clusters to merging ones. 
This makes the CB supercluster especially interesting for studies of the coevolution of clusters in superclusters.

We may ask how large and massive the largest bound structures (collapsing cores of 
superclusters) in the local Universe are. Their sizes and masses are determined
by the mass in clusters and groups, and by the intercluster matter within them.
The mass and radius of the collapsing core at turnaround of supercluster SCl~A2142
with one very massive cluster are  $M \approx 2.3\times~10^{15}h^{-1}M_\odot$ and
$R_T \approx 8$~\Mpc\ \citep{2018A&A...620A.149E}. The total mass in the collapsing core
is determined by the mass of the main cluster, A2142, and by the mass of galaxy groups and gas
in the core.
The mass and radius of the most massive collapsing core in the 
Sloan Great Wall superclusters 
with one very massive cluster are  $M \approx 1.8\times~10^{15}h^{-1}M_\odot$
and $R_T \approx 7.5$~\Mpc\ \citep{2016A&A...595A..70E}.
In the Shapley supercluster, the mass and radius of the collapsing core
with several merging clusters are $M \approx 1.3\times~10^{16}h^{-1}M_\odot$
and $R_T \approx 12.4$~\Mpc\ \citep{2000AJ....120..523R, 2004ogci.conf...71B, 2015A&A...575L..14C}.
The collapsing core of the Shapley supercluster is the largest and most massive 
bound structure in the local Universe found so far, and the core of the CB is the second
most massive. We found that the radius of the core of the CB, which may collapse in the future,
is $R_{\mathrm{FC}} \approx 12.5$~\Mpc, with a mass of $M_{\mathrm{FC}} \approx 4.7\times~10^{15}M_\odot$,
which is lower than that in the Shapley supercluster.

Even the most massive clusters in the local Universe are still growing,
as shown by their merging and infalling structures.
Collapsing cores in the CB supercluster, as well as in
SCl~A2142, are surrounded by the zero-gravity zone that borders the bound and unbound
structures \citep{2015A&A...577A.144T, 2020A&A...641A.172E}.
This suggests  that the supercluster cores themselves have stopped growing. 
The masses and sizes of the most massive possibly collapsing supercluster
cores are determined by the mass of structures within them, first of all,
by the mass and number of the richest clusters. 
The masses of the most massive clusters in the local Universe and in simulations  are of about
$M \approx 10^{15}h^{-1}M_\odot$ \citep{2016A&A...587A.158A, 2015JKAS...48..213K}. 
It would be an interesting task 
to further search 
for high-density cores of superclusters that 
embed several close and/or merging high-mass clusters, 
which could define the most massive bound structures in the Universe.
It could also be interesting to compare the observed number of the most massive
collapsing cores of superclusters with that from simulations. 
For example, \citet{2011MNRAS.413.1311Y} and \citet{2012ApJ...759L...7P}
showed that although the probability of finding in the local Universe as many very rich superclusters
as in the $\Lambda$CDM model is very low, the number of these systems is 
still compatible with the predictions of the $\Lambda$CDM model. 
\citet{2011MNRAS.417.2938S} showed that having both the Shapley supercluster
and the Sloan Great Wall in the local Universe is barely compatible with the
initial Gaussian density fluctuations. The number of massive collapsing cores of
superclusters may add another constraint on the $\Lambda$CDM model and 
on the nature of the primordial fluctuation field.
\citet{2011MNRAS.417.2938S} estimated that the masses of these superclusters are about $M \approx 1.8\times~10^{16}h^{-1}M_\odot$ (the Shapley supercluster)
and $M \approx 1.2\times~10^{17}h^{-1}M_\odot$ (the Sloan Grea Wall).
\citet{2016A&A...595A..70E} estimated using the same supercluster mass estimation method
as used in this paper that the total mass of the Sloan Great Wall is 
$M \approx 2.4\times~10^{16}h^{-1}M_\odot$, which is lower than the \citet{2011MNRAS.417.2938S}
estimate. The total mass of the Corona Borealis supercluster, as found in this paper
($M \approx 4.7\times~10^{15}h^{-1}M_\odot$, mass of the future collapse region),
is approximately 5 to 25 times lower than these estimates. 
The large differences in estimated supercluster masses show that we need
to refine the methods for estimating supercluster masses.  
To strenghten  the predictions by \citet{2011MNRAS.417.2938S}, 
the CB supercluster alone is probably not massive enough in comparison
with extreme superclusters and their complexes, such as the Shapley supercluster and
the Sloan Great Wall. To test the  predictions by \citet{2011MNRAS.417.2938S},
we need to search for  and analyse the properties 
of massive (collapsing cores of) superclusters, especially in the local Universe.
This is an interesting task for future studies that will help us 
to test possible constraints for cosmological models.

\subsection{Why the Corona Borealis?}
\label{sect:cblarge} 

Here we try to answer the question why the CB will probably become
one of the most massive bound systems in the Local Universe.
Galaxy clusters act as attractors that lie at the crossing of galaxy filaments.
In the high-density cores of superclusters, clusters and groups 
are connected by a number of filaments, and therefore 
the number of connections is large and groups are typically richer there than elsewhere
\citep[see also][about the environmental enhancement of galaxy groups near rich 
clusters]{2003A&A...401..851E, 2005A&A...436...17E}. Thus the high-density cores of superclusters
act as great attractors on supercluster scales. 
The CB is located in the dominant supercluster plane,
at the crossing of three very rich supercluster chains
that embed, among other superclusters, the Sloan Great Wall, and the structures of one
of the shell-like structures detected in the nearby Universe
\citep{1997A&AS..123..119E, 2011A&A...532A...5E, 2016A&A...587A.116E}.
This specific location might cause the CB to be so rich and massive,
which will lead to the formation of very massive and large bound systems
in the future.

\citet{2011A&A...531A.149S}
showed that the density waves of different scales affect the formation of galaxy systems
of different richness. Rich galaxy clusters and high-density
cores of superclusters form in regions of high environmental
density, where positive sections of medium- and large-scale 
density perturbations combine. The CB is an example of one of such region.
Moreover,  a $120$~\Mpc\ scale has been found in the cosmic web 
\citep{1990Natur.343..726B, 1994MNRAS.269..301E,
2016A&A...587A.116E}. This scale is determined by the  distribution
of galaxy clusters and superclusters, but not necessarily by the distribution
of the very rich clusters. To understand the formation
of the richest clusters and supercluster cores better, it would be an interesting task on its own 
to search for possible characteristic scales in their distribution and for their 
possibly specific locations in the cosmic web.

\section{Summary}
\label{sect:sum} 

We studied the galaxy content, substructure, and connectivity
of the richest galaxy clusters and their spheres of influence
in the Corona Borealis supercluster, and we
predicted the future evolution of the supercluster. 
We found that the density contrast at the boundaries of the spheres of
influence around the richest clusters in the CB is $\Delta\rho \approx 30$.
This suggests that they passed the turnaround and started to collapse around redshifts
of approximately $z \approx 0.4$. Our main conclusions are listed below.

\begin{itemize}
\item[1)]
Rich clusters in the CB are dynamically active, as shown by their
infalling and possibly merging subclusters, groups, and clusters near them.
\item[2)]
The total connectivity of clusters varies from 2 to 9, and the
number of long filaments that begin near clusters varies from 1 to 5. 
\item[3)]
The values of the cluster radii, $R_{\mathrm{cl}}$ 
(the radius of a cluster for a one-component cluster, 
and the radius of the main component
in clusters with several components, usually called splashback radius)
are $R_{\mathrm{cl}} \approx 2 - 3$ ~$R_{\mathrm{vir}}$.  
At clustercentric distances $D_c \simeq R_{\mathrm{cl}}$, 
there is a small maximum in the distribution of galaxies. 
\item[4)]
Around each cluster lies a sphere of influence with radius $R_{\mathrm{30}}$
that corresponds to the density contrast around the cluster,
$\Delta\rho \approx 30$. 
Within this sphere, all galaxies and groups are falling into clusters.
The high-density contrast shows that within the spheres of influence, galaxy systems
are collapsing, and the collapse began at redshifts of approximately $z \approx 0.3 - 0.4$.
\item[5)]
Galaxy content in clusters and in their spheres of influence varies strongly.  
In the main part of the CB in the most massive cluster, A2065, and in
its sphere of influence 
the fraction of galaxies with very old stellar populations is lower than 
in the other two clusters in this part of the CB. Cluster 
Gr2064 has a high fraction of star-forming galaxies in an infalling
group in the turnaround region of the cluster. Star formation in galaxies in this group
may be triggered by infall and collapse. 
\item[6)]
Galaxies in transformations, red star-forming galaxies and recently quenched galaxies,
lie in the infall zones of clusters and in infalling groups.
\item[7)]
The galaxy populations in clusters A2142 and in A2061 are statistically similar.
In these clusters, the galaxies have older stellar populations than in A2065 and A2089.
\item[8)]
During the future evolution, two parts of the CB will separate.
The main part forms a collapsing region that will probably become one of the largest and most massive
bound systems in the nearby Universe.
\end{itemize}

We found that the density contrast at the boundaries of the spheres of
influence around the richest clusters in the CB is $\Delta\rho \approx 30$.
To understand the formation and evolution of rich galaxy clusters,
we need to continue to study the environment of clusters to 
determine whether the characteristic density contrast $\Delta\rho = 30$ 
can be found around other clusters, and what the implications of this are
for the understanding of cluster evolution.
This finding suggests that they have passed turnaround and started to collapse around redshifts
of approximately $z \approx 0.4$. Simulations show that at redshift $z = 0.5,$  
galaxy clusters have half of their current mass \citep{2015JKAS...48..213K}.
This means that studies of galaxy  clusters and their environment
at redshifts between $z = 0 - 0.5$ are especially important 
for understanding the evolution of galaxy clusters and galaxies within them.
These data will be provided by future surveys, for example, 
the 4MOST cluster survey 
and 4-MOST Cosmology Redshift Survey (CRS)
\citep{2019Msngr.175....3D, 2019Msngr.175...39F, 2019Msngr.175...50R}.
The CRS survey can be extended  with HI data from the SKA survey \citep{2017MNRAS.470.3220W}.

The massive and large collapsing supercluster cores are the largest objects in the Universe
to collapse now or in the future. It would be an interesting task on its own
to search for similar structures from observations, and to compare the number
and properties of these structures in observations and simulations.
With data from the next generation of large galaxy redshift surveys, for instance, Euclid
\citep{2011arXiv1110.3193L}, DESI 
\citep{2013arXiv1308.0847L, 2016arXiv161100036D}, and  J-PAS \citep{2014arXiv1403.5237B},
it will become possible to study such systems over a reasonable span of the history
of the Universe, enabling a direct comparison to the evolution seen in large-scale
cosmological simulations. This comparison will serve as a test for cosmological
models and may give valuable information about dark matter and dark energy.

\begin{acknowledgements}
We thank the referee for valuable comments and suggestions
which helped us to improve the paper.
We thank Changbom Park, Boris Deshev, Gayoung Chon, and Mirt Gramann 
for useful discussions.
We are pleased to thank the SDSS Team for the publicly available data
releases.  Funding for the Sloan Digital Sky Survey (SDSS) and SDSS-II has been
provided by the Alfred P. Sloan Foundation, the Participating Institutions,
the National Science Foundation, the U.S.  Department of Energy, the
National Aeronautics and Space Administration, the Japanese Monbukagakusho,
and the Max Planck Society, and the Higher Education Funding Council for
England.  The SDSS website is \texttt{http://www.sdss.org/}.
The SDSS is managed by the Astrophysical Research Consortium (ARC) for the
Participating Institutions.  The Participating Institutions are the American
Museum of Natural History, Astrophysical Institute Potsdam, University of
Basel, University of Cambridge, Case Western Reserve University, The
University of Chicago, Drexel University, Fermilab, the Institute for
Advanced Study, the Japan Participation Group, The Johns Hopkins University,
the Joint Institute for Nuclear Astrophysics, the Kavli Institute for
Particle Astrophysics and Cosmology, the Korean Scientist Group, the Chinese
Academy of Sciences (LAMOST), Los Alamos National Laboratory, the
Max-Planck-Institute for Astronomy (MPIA), the Max-Planck-Institute for
Astrophysics (MPA), New Mexico State University, Ohio State University,
University of Pittsburgh, University of Portsmouth, Princeton University,
the United States Naval Observatory, and the University of Washington.

The present study was supported by the ETAG projects 
PRG1006,  PUT1627, PUTJD907,  and by the European Structural Funds
grant for the Centre of Excellence "Dark Matter in (Astro)particle Physics and
Cosmology" (TK133).
This work has also been supported by
ICRAnet through a professorship for Jaan Einasto.
We applied in this study R statistical environment 
\citep{ig96}.

\end{acknowledgements}

\bibliographystyle{aa}
\bibliography{cb0.bib}

\begin{appendix}

\section{Coordinate system used in acceleration field calculations.}
\label{sect:coord} 

\begin{figure}[ht]
\centering
\resizebox{0.44\textwidth}{!}{\includegraphics[angle=0]{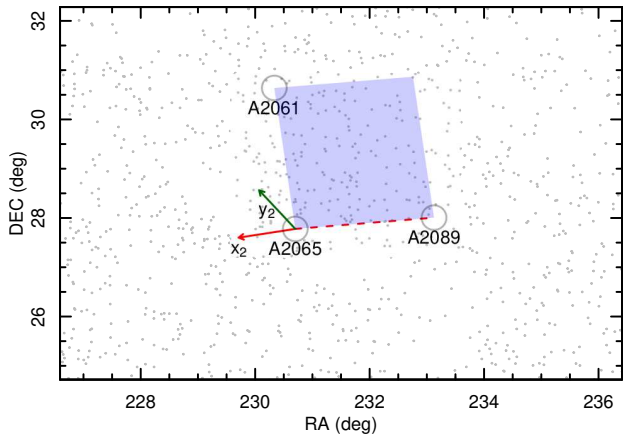}}
\caption{
Coordinate system used in calculating the acceleration field
of clusters in the plane of the sky. The blue plane shows the plane of 
the three clusters that define the coordinate system.
The $x_2$ coordinate direction connects A2089 and A2065, the $y_2$ coordinate
is  perpendicular to $x_2$  
and is in the same plane as determined by the clusters. The
$z_2$ coordinate is perpendicular to both
$x_2$ and $y_2$. 
}
\label{fig:coord}
\end{figure}

To calculate the acceleration field for the main part of the CB supercluster,
we assumed that
the dominant acceleration is caused by the most massive clusters in it
(A2065, A2061, and A2089). In
order to show their effect most strongly, we used in the calculations
a coordinate system that
is located in the plane determined by these three clusters. 
The
$x_2$ coordinate is determined by connecting A2089 and A2065, the $y_2$ coordinate
is defined to be perpendicular to $x_2$  
and to be in the same plane as determined by these three
clusters. The
$z_2$ coordinate is perpendicular to both
$x_2$ and $y_2$. 
The zero point of this coordinate system is the mid-point between clusters.
This coordinate system is designed to keep the majority of the acceleration field in
the plane of the figures, hence to show how the system may evolve in the clearest manner. 
This coordinate system is shown in Fig.~\ref{fig:coord}.

\end{appendix}

\end{document}